\documentclass{raa}            

\usepackage{graphicx, times}             
\usepackage{natbib}
\usepackage{amssymb, amsmath}
\bibpunct{(}{)}{;}{a}{}{, }

\usepackage{multirow}
\usepackage[pagebackref=true]{hyperref}
\begin{document}

\title{Enhancing cosmological constraints
with nonlinear tanh transformations of Hermite-Gaussian Derivative
fields}

   \volnopage{Vol.0 (20xx) No.0,  000--000}      
   \setcounter{page}{1}          

\author{Zhiwei Min
      \inst{1}
    \and Ye Ma
      \inst{1}
      \and Zhujun Jiang
      \inst{1}
      \and Jiacheng Ding
      \inst{3}
   \and Fenfen Yin
      \inst{2}
   \and Le Zhang
      \inst{1,5}
    \and Xiao-Dong Li
      \inst{1,4,5}
   }

\institute{School of Physics and Astronomy, Sun Yat-Sen University, Zhuhai 519082, China; {\it zhangle7@mail.sysu.edu.cn}; {\it lixiaod25@mail.sysu.edu.cn}\\
\and
School of Physics and Electronic Information Engineering, Tongren University, Tongren 554300, China; {\it dsjyff@gztrc.edu.cn}\\
\and
Shanghai Astronomical Observatory, Chinese Academy of Sciences, Shanghai 200030, P. R. China; {\it dingjch@shao.ac.cn}\\
\and
Peng Cheng Laboratory, No. 2, Xingke 1st Street, Shenzhen 518000, China;\\
\and
CSST Science Center for the Guangdong--Hong Kong--Macau Greater Bay Area, SYSU, Zhuhai 519082, China\\
\vs \no
   {\small Received 20XX Month Day; accepted 20XX Month Day}
}

\abstract
{A key goal in large-scale structure analysis is to extract multi-scale information to improve cosmological parameter constraints. In particular, higher-order derivative
fields are especially valuable as they capture the geometric and topological information of
the cosmic web that is highly sensitive to cosmological parameters. Traditional derivative-based methods, such as finite-difference or Fourier approaches, suffer from noise amplification at small scales and cannot stably capture multi-scale features. We present a robust two-step framework: first, stable multi-scale arbitrary-order derivatives are obtained via Hermite–Gaussian convolutional filters that suppress small-scale noise; second, a $\tanh$ nonlinear transformation compresses extreme density contrasts and enhances the visibility of cosmic web structures. Using the Quijote simulations, we show that combining multi-scale first-order spectra yields improvements of 1.2-3.0$\times$ across all seven cosmological parameters, while multi-order spectra at a fixed scale provide 1.3-2.9$\times$ gains. The most comprehensive combination achieves nominal gains of 2.0-5.3$\times$. Our method offers a robust approach to extracting additional cosmological information for future surveys.
}

   \authorrunning{Min et al.}            
   \titlerunning{Cosmological constraints from derivative
   	fields}  

   \maketitle

%
%
\section{Introduction}           
\label{sect:intro}

The large-scale structure (LSS) of the Universe directly reflects its evolution and composition \citep{Peebles1980}. Different cosmological models predict distinct structural patterns as gravity amplifies initial density fluctuations. This process forms the "cosmic web"—a network of clusters (nodes),  filaments,  and voids \citep{Bond1996}. Key properties of this web,  like the abundance of clusters and connectivity of filaments,  are highly sensitive to fundamental cosmological parameters \citep{Goh2018}. Therefore,  measuring the cosmic web's geometry provides a powerful way to test cosmological theories. A central challenge is efficiently extracting this information from modern astronomical surveys.

The standard two-point statistics—the correlation function and power spectrum—have successfully constrained cosmology from galaxy surveys \citep{Schneider_2006,BOSS:2016wmc,Planck:2018vyg}. However,  they capture only Gaussian information. The late-time cosmic web,  shaped by non-linear gravity,  is rich in non-Gaussian information \citep{Scoccimarro1999, Sefusatti2005}. This information,  encoding structure morphology,  is crucial for breaking parameter degeneracies \citep{Heavens2007}. Upcoming fourth generation surveys (e.g.,  DESI,  LSST,  Euclid, CSST) \citep{LSST2017, Euclid2020,Gong:2025ecr,DESI:2025zgx} will provide unprecedented data on non-linear scales. To leverage this,  we must develop new statistical tools beyond the power spectrum.

Derivative fields of the cosmic density field offer a promising avenue for probing the structural details of the cosmic web. The underlying physical intuition is twofold. First, the geometry of cosmological structures is imprinted in how the density field changes in space. First‑order derivatives (the gradient field) indicate the direction and magnitude of the steepest density variations, highlighting the boundaries of voids and sheets. Second‑order derivatives, encoded in the Hessian matrix, quantify the local curvature of the density field \citep{Hahn2007, ForeroRomero2009}. The eigenvalues of the Hessian can be used to classify the local environment into nodes, filaments, sheets, or voids, providing a direct measure of the cosmic web’s morphology \citep{Aragon-Calvo:2008ghu}. Second, the derivative fields of the galaxy density field are tightly linked to the velocity field via gravitational dynamics. In the linear regime, the velocity field is related to the density‑field gradient through the Poisson equation \citep{Kaiser1987}, and it is particularly sensitive to key cosmological parameters—most notably the growth rate $f(z) = \Omega_m(z)^{0.55}$ and the amplitude of fluctuations $\sigma_8$. Because the velocity field responds directly to the underlying gravitational potential, its relation to the density gradient provides a cleaner, more linear tracer of cosmological parameters than the density field alone. Consequently, derivative‑based diagnostics that connect density gradients to velocity information are expected to retain enhanced cosmological sensitivity, especially for parameters that govern structure growth and expansion history \citep{Strauss:1995fz,Johnston2012ReconstructingTV}. Therefore,  statistics derived from these derivative fields have the potential to be exceptionally sensitive to cosmological parameters that influence the growth and shape of structures. Previous studies have indeed explored using Minkowski functionals \citep{Schmalzing1997},  the Hessian for web classification \citep{Hahn2007, ForeroRomero2009},  or wavelet transforms to capture similar information \citep{Mallat2012, Regaldo2023, Jiang:2025wle} at different scales.

However, existing derivative-based methods suffer from a fundamental trade-off between simplicity and stability. Direct approaches like finite-differencing or Fourier-based derivatives are inherently single-scale and cannot naturally extract information across multiple scales—a critical limitation for cosmological parameter constraints where different scales carry complementary information \citep{Ivanov2022}. Techniques like finite-differencing are easy to implement but amplify small-scale noise, making them unsuitable for higher-order derivatives which are essential for capturing complex patterns. While Fourier-based methods can compute derivatives exactly, they require careful, often arbitrary, filtering to suppress noise, and lack a principled way to combine the information from multiple derivative orders. Furthermore, when applied directly to discrete point clouds (e.g., galaxy or halo distributions), computing derivatives necessitates estimating the local density field and its gradients for each particle, typically through computationally intensive algorithms like K-nearest neighbors (KNN) \citep{banerjee2021nearest,gangopadhyay2025geometric}. This process becomes prohibitively expensive for high-order derivatives across large volumes. The absence of a robust and computationally efficient framework for multi-scale derivative analysis has therefore been a major obstacle in maximizing cosmological information extraction.

In this work,  we introduce a novel method based on Hermite polynomials-weighted Gaussian convolution to efficiently generate derivative fields of the density field. A key advantage of our approach is that convolving the density field with a Hermite-Gaussian kernel is mathematically equivalent to first smoothing the field with a Gaussian kernel and then applying the derivative operator. This process automatically imposes an exponential cutoff $\exp(-k^2)$, in Fourier space,  which suppresses noisy,  high-$k$ modes and acts as a natural regularizer—a feature not present in simple finite-difference derivative schemes. Importantly, by varying the smoothing scale of the convolutional kernel, our Hermite-convolution approach automatically extracts information across
multiple scales. We focus on fields up to second order to capture information from gradients,  curvature,  and more complex morphology.  To enhance the signal-to-noise ratio for practical applications,  we apply a non-linear transformation,  the hyperbolic tangent function($\tanh$) to the derivative fields in configuration space. This nonlinear transformation compresses extreme density contrasts, thereby enhancing the visibility of cosmic web structures. Consequently, the derivative fields are converted into morphological statistics that exhibit higher sensitivity to cosmological geometry. \citep{Xiao2022} extend traditional mark-weighted statistics by using powers of the normalized density field gradient, $(|\nabla\rho|/\rho)^\alpha$, as the weight. This operation produces an effect on the density field that is similar to a $\tanh$ transformation. Our analysis demonstrates that the tanh-transformed fields yield improved cosmological parameter constraints, particularly for structure-sensitive parameters $\Omega_b$, $h$, $M_\nu$, and $w$, as will be shown in the following sections.

This paper is organized as follows. In Section 2,  we describe the Quijote suite of N-body simulations,  which serve as the data base for our analysis. Section 3 details our methodology,  including the mathematical formalism of Hermite convolution for calculating derivative fields,  the computation of power spectra from these fields,  and the Fisher matrix formalism used to quantify the constraining power on cosmological parameters. Our main results,  presenting a comparison of the information content in convolutional fields of different orders and a comparison with the standard matter power spectrum,  are shown in Section 4. Finally,  we summarize our findings and discuss their implications for future cosmological analyses in Section 5.


\section{Data}
This section describes the Quijote simulation suite used in our analysis and the necessary data processing steps performed to match real observational conditions. We utilize these simulations to compute the covariance matrices and numerical derivatives  required for constraining cosmological parameters within the Fisher matrix framework.

\label{sect:data}
\subsection{Simulations}\label{sec:simu}
The Quijote simulation suite\citep{Villaescusa-Navarro:2019bje} comprises over 43, 000 full $N$-body simulations spanning a range of cosmological parameters. These simulations are executed with the TreePM code \textsc{Gadget}-III,  an advanced version of the publicly available \textsc{Gadget}-II\citep{Springel2005TheCS},  starting from initial conditions (ICs) set at redshift $z = 127$. For most cosmologies,  the ICs are generated using second-order Lagrangian perturbation theory (2LPT). However,  for massive neutrino cosmologies  (see Tab.~\ref{tab:cosmo_params}),  the Zel'dovich approximation is employed instead. Additionally,  a subset of simulations in the fiducial cosmology also use Zel'dovich ICs to ensure consistent numerical derivative calculations for neutrino mass $M_\nu$. The fiducial cosmology is defined by $(\Omega_m,  \Omega_b,  h,  n_s,  \sigma_8,  M_\nu,  w) = (0.3175,  0.049,  0.6711,  0.9624,  0.834,  0~\mathrm{eV},  -1)$.

Halos in the Quijote simulations are identified from the cold dark matter distribution using a Friends-of-Friends (FoF) algorithm with a linking length of $b = 0.2$. We maintain a fixed halo number density of $n_{\mathrm{halo}} = 2 \times 10^{-4} \,  h^{3} \,  \mathrm{Mpc}^{-3}$ (see details in Section~\ref{sec:data_preprocessing}). This choice results in a minimum halo mass cutoff that varies with the cosmological parameter.

In this work,  we utilize the FoF dark matter halo catalogs at redshift $z = 0.5$. We compute covariance matrices from 10, 000 realizations of the fiducial cosmology and numerical derivatives from 500 simulations  where individual parameters are varied above or below their fiducial values. All simulations contain $512^3$ cold dark matter particles (plus $512^3$ neutrino particles for massive neutrino cases) within a $(1h^{-1}\mathrm{Gpc})^3$ box. The cosmological simulations used for covariance and derivative calculations are summarized in Tab.~\ref{tab:cosmo_params}.

\begin{table}[htbp]
\centering
\caption{Description of the N-body simulations used in the Fisher analysis. $\Omega_m$ is the matter density parameter,  $\Omega_b$ is the
baryon density parameter,  $h$ is the dimensionless Hubble constant, $n_s$ is the spectral index, $\sigma_8$ is the root-mean-square amplitude
of the linear matter fluctuations at 8 $h^{-1}\mathrm{Mpc}$,   $M_\nu$is the sum of neutrino masses, and $w$ is the dark energy equation of state parameter.}
\label{tab:cosmo_params}
\begin{tabular}{|c|ccccccccc|}
\hline
Name & $\Omega_{m}$ & $\Omega_{b}$ & $h$ & $n_{s}$ & $\sigma_{8}$ & $M_{\nu}$ (eV) & $w$ &ICs&realizations\\
\hline
\hline
Fid     & \underline{0.3175} & \underline{0.049}  & \underline{0.6711} & \underline{0.9624} & \underline{0.834} & \underline{0}  & \underline{$-1$} &2LPT/Zel’dovich&10, 000/500\\
\hline
$\Omega_{m}^{+}$  & \underline{0.3275} & 0.049  & 0.6711 & 0.9624 & 0.834 & 0 & $-1$ &2LPT&500\\
\hline
$\Omega_{m}^{-}$  & \underline{0.3075} & 0.049  & 0.6711 & 0.9624 & 0.834 & 0 & $-1$ &2LPT&500\\
\hline

$\Omega_{b}^{+}$  & 0.3175 & \underline{0.050}  & 0.6711 & 0.9624 & 0.834 & 0 & $-1$ &2LPT&500\\
\hline
$\Omega_{b}^{-}$  & 0.3175 & \underline{0.048}  & 0.6711 & 0.9624 & 0.834 & 0 & $-1$ &2LPT&500\\
\hline

$h^{+}$ & 0.3175 & 0.049  & \underline{0.6911} & 0.9624 & 0.834 & 0 & $-1$ &2LPT&500\\
\hline
$h^{-}$ & 0.3175 & 0.049  & \underline{0.6511} & 0.9624 & 0.834 & 0 & $-1$ &2LPT&500\\
\hline
$n_{s}^{+}$ & 0.3175 & 0.049  & 0.6711 & \underline{0.9824} & 0.834 & 0 & $-1$ &2LPT&500\\
\hline
$n_{s}^{-}$ & 0.3175 & 0.049  & 0.6711 & \underline{0.9424} & 0.834 & 0 & $-1$ &2LPT&500\\
\hline
$\sigma_{8}^{+}$ & 0.3175 & 0.049  & 0.6711 & 0.9624 & \underline{0.849} & 0 & $-1$ &2LPT&500\\
\hline
$\sigma_{8}^{-}$ & 0.3175 & 0.049  & 0.6711 & 0.9624 & \underline{0.819} & 0 & $-1$ &2LPT&500\\
\hline

$M_{\nu}^{++}$ & 0.3175 & 0.049  & 0.6711 & 0.9624 & 0.834 & \underline{0.2} & $-1$ &Zel’dovich&500\\
\hline
$M_{\nu}^{+}$ & 0.3175 & 0.049  & 0.6711 & 0.9624 & 0.834 & \underline{0.1} & $-1$ &Zel’dovich&500\\
\hline
$w^{+}$ & 0.3175 & 0.049  & 0.6711 & 0.9624 & 0.834 & 0 & \underline{$-1.05$} &Zel’dovich&500\\
\hline
$w^{-}$ & 0.3175 & 0.049  & 0.6711 & 0.9624 & 0.834 & 0 & \underline{$-0.95$} &Zel’dovich&500\\
\hline
\end{tabular}
\end{table}

\subsection{Data preprocessing}\label{sec:data_preprocessing}
To simulate key observational effects,  we preprocessed the halo position catalogs from the Quijote simulations.

1) Redshift-space distortions (RSD): We introduced RSD along the line-of-sight (LoS) using:
\begin{equation}\label{eq:rsd}
\bm{s}=\bm{r}+\frac{\bm{v} \cdot \hat{z}}{a H(a)}\hat{z}, 
\end{equation}
where $\bm{r}$,  $\bm{s}$ are the position of halos in real space and redshift space respectively. $\hat{z}$ is the unit vector along LoS,   $\bm{v}$ is the peculiar velocity of halos,  and $H(a)$ is the Hubble parameter at scale factor $a$. 

2) Fiducial cosmology conversion: Observational analyses assume a fixed fiducial cosmology to convert redshifts to distances. We aligned our mocks with the Planck 2018 measurements\citep{Planck:2018vyg},  where $\Omega_m = 0.3071$,  $w = -1$. The relation between the Quijote cosmologies and the fiducial cosmology is expressed by:
\begin{equation}\label{eq:fiducial}
s_\perp = s_\perp^0\frac{d^{f}_A(z)}{d_A(z)}\, , ~\quad
s_\parallel = s_\parallel^0\frac{H(z)}{H^{f}(z)}
\end{equation}
where $d_A (z)$ and $H(z)$ represent the angular diameter distance and the Hubble parameter at redshift $z$,  respectively.  The variables $s_\parallel^0$ and $s_\perp^0$ represent the comoving coordinates in each Quijote simulation. The superscript $f$ denotes the fiducial cosmology,  while $\bm{s}_\perp$ and $\bm{s}_\parallel$ represent components perpendicular and parallel to LoS,  respectively.  The objective of this transformation process is to more accurately reflect the data analyzed in real observations \citep{BOSS:2016wmc, Min:2024dgd}.

3) Box cutting: After conversion,  the
box sizes are no longer the same in all three dimensions. We cut them to cubes of side length $L = 744~h^{-1}{\rm Mpc}$ for a uniform analysis. Consequently,  only the halos within such box in each simulation are considered.

4) Mass cutoff: Considering that dark matter (DM) halos with very low mass contribute significant noise,  we implemented a cutoff for small mass halos. This cutoff was chosen appropriately such that the number of DM halos has a density equal to $n_\text{halo} = 2 \times 10^{-4}~h^3{\rm Mpc}^{-3}$ in each box to be compatible with  spectral observations. 

\section{Methodology}
This section describes the Hermite convolutional derivative filters and presents the resulting Hermite convolutional density fields along with their tanh-transformed counterparts. We then analyze the power spectra of these fields and examine the effects of both the convolution and tanh operations. Finally, we introduce the Fisher formalism components used to constrain cosmological parameters, including the covariance matrix and numerical derivatives with respect to different parameters.
\subsection{Hermite Convolution Derivative Filter}
To extract  derivative information of the cosmic density field,  we employ a Hermite convolution based derivative filter. This approach allows us to compute arbitrary order partial derivatives of the smoothed density field.

\begin{figure}
    \centering
    \includegraphics[width=0.7\linewidth]{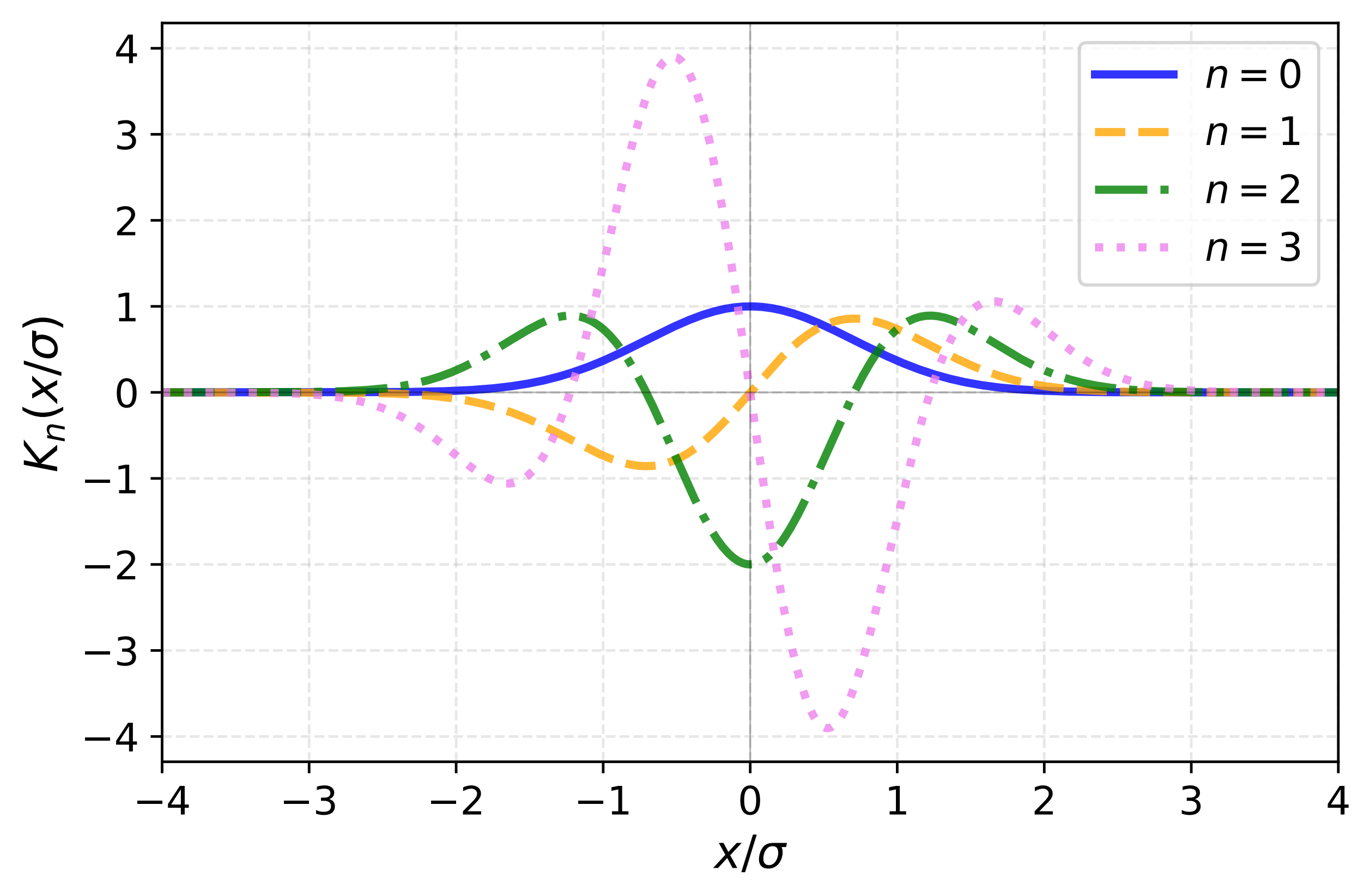}
    \caption{The functional forms of $K_n(x/\sigma) = H_n(x/\sigma)\exp(-x^2/\sigma^2)$ for different orders $n = (0, 1, 2, 3)$. These kernel functions exhibit distinct oscillatory and smoothing characteristics across the normalized coordinate $x/\sigma$. The $n = 0$ case (blue solid line) shows a smooth,  non-oscillatory Gaussian profile,  while higher-order kernels($n\ge 1$), display increasing numbers of zero-crossings and oscillatory behavior,  which is characteristic of Hermite polynomial-based filters. These kernels are generalized to 3D and used in our Hermite convolution framework to extract cosmological information from the density field. }
    \label{fig:hermite-gaussian-kernel}
\end{figure}

The $n$-th order Hermite polynomials (physicist’s form) are a set of orthogonal polynomials defined by:
\begin{equation}
    \label{eq:hermite poly}
    H_n(\xi) = (-1)^n\exp(\xi^2) \frac{d^n}{d\xi^n}\exp(-\xi^2) 
\end{equation}
The explicit expressions of the first 4 Hermite polynomials are:
\begin{equation}
    \label{eq:hermite_poly_explicit}
    \begin{aligned}
        H_0(\xi) &= 1, \\
        H_1(\xi) &= 2\xi, \\
        H_2(\xi) &= 4\xi^2 - 2, \\
        H_3(\xi) &= 8\xi^3 - 12\xi.
    \end{aligned}
\end{equation}
They satisfy the orthogonality relation:
\begin{equation}
    \label{eq:ortho}
    \int_{-\infty}^{\infty}H_n(\xi)H_m(\xi)\exp(-\xi^2)d\xi = \sqrt{\pi}2^nn!\delta_{mn}.
\end{equation}
In three dimensions,  the functions formed by Hermite polynomials multiplied by a Gaussian, constitute a complete orthogonal basis for the space of square-integrable functions. This property makes them exceptionally well-suited for constructing multi-scale convolution kernels,  in a manner analogous to a wavelet scattering transform \citep{Regaldo2023, Jiang:2025wle}. Fig.~\ref{fig:hermite-gaussian-kernel} shows the one-dimensional Hermite-Gaussian kernels for various orders. Kernels of order $n \ge 1$ exhibit oscillatory behavior,  enabling them to capture density gradients and asymmetric environmental features. This property is particularly advantageous for identifying regions with sharp environmental transitions,  such as the boundaries of filaments,  voids,  and clusters \citep{Hahn2007, ForeroRomero2009}.

According to Eq. \ref{eq:hermite poly}, a  key identity relates the product of Hermite polynomials and a Gaussian to derivatives of the Gaussian:

\begin{equation}
H_{n_{x}}\left(\frac{x}{\sigma}\right) H_{n_{y}}\left(\frac{y}{\sigma}\right) H_{n_{z}}\left(\frac{z}{\sigma}\right) e^{-r^{2}/\sigma^{2}}
= (-1)^{n}\sigma^{n}\frac{\partial^{n}}{\partial x^{n_{x}}\partial y^{n_{y}}\partial z^{n_{z}}}e^{-r^{2}/\sigma^{2}}, 
\end{equation}
where $n = n_{x} + n_{y} + n_{z}$, and $r^2 = x^2+y^2+z^2$.

Convolving both sides with the dark matter halo overdensity field $\delta(\mathbf{x})$ and using the commutativity of convolution with differentiation,  we obtain the filter identity:

\begin{equation}
\left[H_{n_{x}}\left(\frac{x}{\sigma}\right)H_{n_{y}}\left(\frac{y}{\sigma}\right)H_{n_{z}}\left(\frac{z}{\sigma}\right)e^{-r^{2}/\sigma^{2}}\right] \ast \delta
= (-1)^{n}\sigma^{n}\frac{\partial^{n}}{\partial x^{n_{x}}\partial y^{n_{y}}\partial z^{n_{z}}}\delta_{\sigma}, 
\end{equation}
where the star symbol $\ast$ stands for the convolution operation, and $\delta_{\sigma} = G \ast \delta$ is the Gaussian-smoothed density field with $G(\mathbf{x}) = e^{-r^{2}/\sigma^{2}}$.

This identity establishes that convolving the density field with a Hermite–Gaussian kernel directly yields the corresponding partial derivatives of the smoothed density field. The scale parameter $\sigma$ controls the smoothing scale,  while the polynomial orders  $(n_x,  n_y,  n_z)$ determine the type and order of differentiation.

\subsection{Convolutional density fields}
In this work,  we choose to work in Fourier space rather than configuration space due to the lower computational cost. Specifically,  the normalized convolutional field is
\begin{equation}
\delta^\sigma_{n_x, n_y, n_z}(\vec{x}) = \frac{1}{(-1)^n\sigma^n}\text{FFT}^{-1}\left[\text{FFT}(K_\sigma) \cdot \text{FFT}(\delta(\vec{x}))\right], 
\end{equation}
where FFT and $\text{FFT}^{-1}$ denote the forward and inverse fast Fourier transform operations, respectively. And $K_\sigma$ is the Hermite-Gaussian convolutional kernel with smoothing scale $\sigma$,  defined as $K_\sigma(x, y, z) = H_{n_{x}}\left(\frac{x}{\sigma}\right) H_{n_{y}}\left(\frac{y}{\sigma}\right) H_{n_{z}}\left(\frac{z}{\sigma}\right) e^{-r^{2}/\sigma^{2}}$. In practice,  we choose $\sigma = 6, 10, 15 h^{-1}\mathrm{Mpc}$ to extract multi-scale information from large scale structure. In principle,  small smoothing scales allow for a more detailed estimation of the local density,  whereas very large scales smooth overdensities to near zero. However,  the smoothing scale $\sigma$ cannot be arbitrarily small. This is limited not only by the need to control the maximum wavenumber in the analysis but also by the resolution of the N-body simulations (the initial condition grid size is $2 h^{-1}\mathrm{Mpc}$) and the sparsity of halos. An excessively small $\sigma$ would cause the estimate of $\delta^\sigma_{n_x, n_y, n_z}(\vec{x})$ to reflect Poisson noise rather than the true derivative signal.

While differentiation is a linear operation in Fourier space and preserves Fisher information in principle, practical observational effects—including noise, survey masks, wavenumber ($k$)‑limit cutoffs, and non‑Gaussianities—can significantly alter this ideal scenario. To enhance the robustness and information extraction capability of our convolutional fields under realistic conditions, we first construct the \textit{magnitude} for each derivative order. The first-order magnitude field $\delta^{\sigma}_1(\vec{x})$ is defined as the norm of the three first-order components with indices $(n_x,n_y,n_z) = (1,0,0), (0,1,0)$, and $(0,0,1)$. Similarly, the second-order magnitude field $\delta^{\sigma}_2(\vec{x})$ is constructed from the components with indices $(n_x,n_y,n_z) = (2,0,0), (0,2,0)$, and $(0,0,2)$, while the mixed second-order magnitude field $\delta^{\sigma}_{2,\mathrm{mix}}(\vec{x})$ uses the components with indices $(n_x,n_y,n_z) = (1,1,0), (1,0,1)$, and $(0,1,1)$:

\begin{equation}\label{eq:norm}
    \begin{split}
        \delta^{\sigma}_1(\vec{x}) &= \sqrt{[\delta^{\sigma}_{1,0,0}(\vec{x})]^2 + [\delta^{\sigma}_{0,1,0}(\vec{x})]^2 + [\delta^{\sigma}_{0,0,1}(\vec{x})]^2},\\
        \delta^{\sigma}_2(\vec{x}) &= \sqrt{[\delta^{\sigma}_{2,0,0}(\vec{x})]^2 + [\delta^{\sigma}_{0,2,0}(\vec{x})]^2 + [\delta^{\sigma}_{0,0,2}(\vec{x})]^2},\\
        \delta^{\sigma}_{2,\text{mix}}(\vec{x}) &= \sqrt{[\delta^{\sigma}_{1,1,0}(\vec{x})]^2 + [\delta^{\sigma}_{1,0,1}(\vec{x})]^2 + [\delta^{\sigma}_{0,1,1}(\vec{x})]^2}.
    \end{split}
\end{equation}

We then apply a nonlinear $\tanh$ transformation to each magnitude field:

\begin{equation}
\label{eq:tanh}
\widetilde{\delta}^\sigma_{J}(\vec{x}) = \tanh\left(\frac{\delta^\sigma_{J}(\vec{x})}{\alpha}\right),
\end{equation}

where $\delta_J^\sigma$ denotes one of the three magnitude fields defined in Eq. \ref{eq:norm}, and $\alpha$ is an adjustable parameter. We set $\alpha = 600$ for the first-order magnitude field $\delta^{\sigma}_1(\vec{x})$, $\alpha = 90$ for the second-order magnitude field $\delta^{\sigma}_2(\vec{x})$, and $\alpha = 50$ for the mixed second-order magnitude field $\delta^{\sigma}_{2,\text{mix}}(\vec{x})$. The value of $\alpha$ is chosen to be approximately the standard deviation of the pixel values of the convolutional fields, ensuring that most field values undergo near-linear mapping while extreme values are compressed towards $ 1$.

\begin{figure}[h!]
    \centering
    \includegraphics[width=0.7\linewidth]{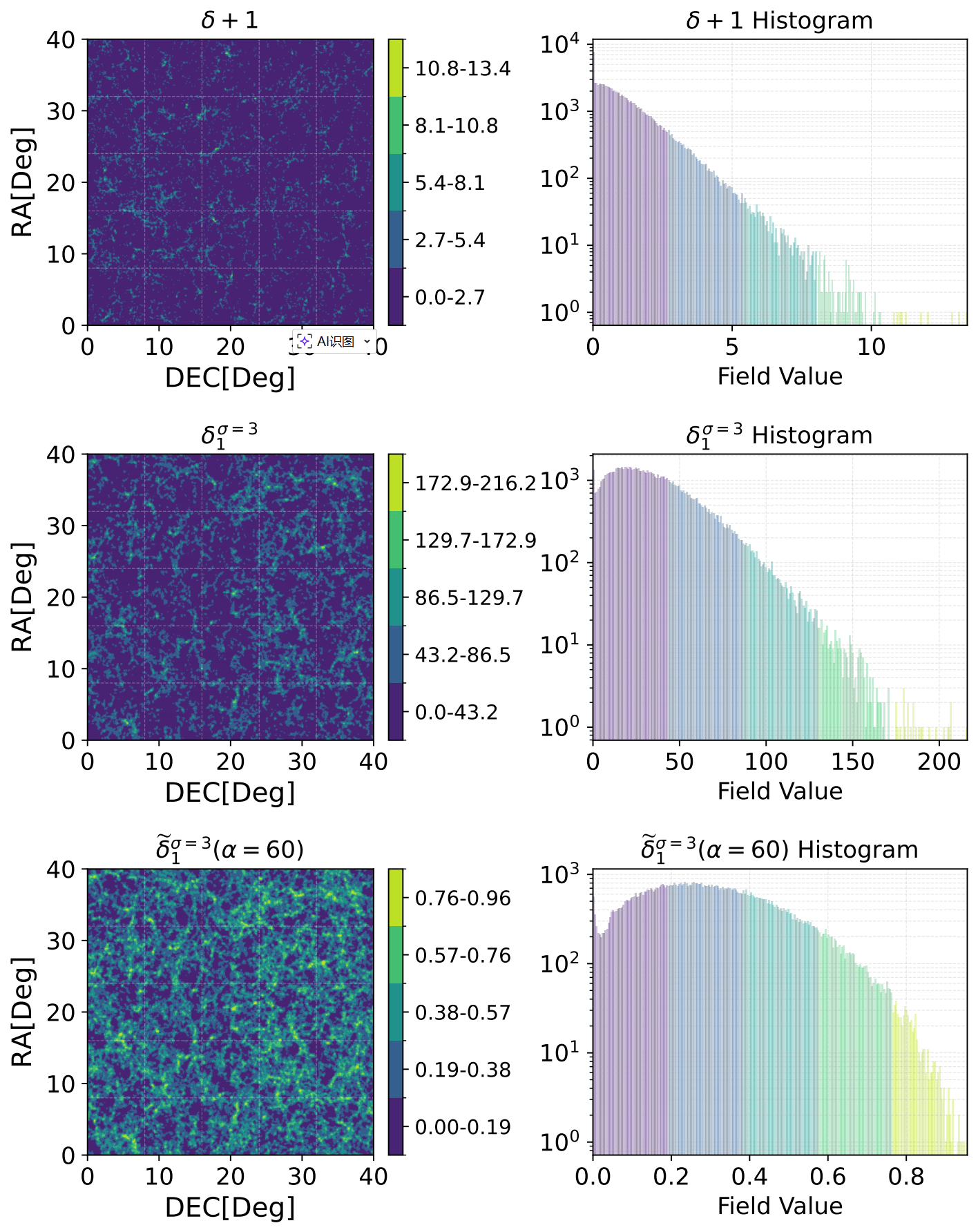}
    \caption{
         Projected halo density fields from the high-resolution \textsc{AbacusSummit} simulation over the redshift slice  $0.475\le z \le 0.538$ ,covering a sky area of $40^\circ\times 40^\circ$. 
        Top row: Raw halo density field $\delta+1$. 
        Middle row: First-order Hermite-Gaussian (HG) convolutional field $\delta^{3}_{1}$ ($\sigma = 3\ h^{-1}\mathrm{Mpc}$). 
        Bottom row: tanh-transformed HG (HG-tanh) field $\widetilde{\delta}^{3}_{1}$ ($\alpha = 60$). 
        The right column shows the corresponding field value histograms. 
        The HG operation acts as a spatial differentiator that enhances structural gradients, sharpening filaments and cluster boundaries. 
        The subsequent tanh transformation compresses extreme values, further improving the visibility of intermediate-density structures and the connectivity of the cosmic web.
    }
    \label{fig:field_compare}
\end{figure}

To clearly demonstrate the effects of the Hermite-Gaussian convolution and the subsequent tanh operation, we present in Figure~\ref{fig:field_compare} the projected density field and its corresponding histograms from the high-resolution \textsc{AbacusSummit} simulation \citep{Maksimova:2021ynf}. The top row displays the raw halo field  $\delta+1$ along with its histogram.
 The middle row shows the Hermite-Gaussian (HG) field $\delta_J^\sigma$, where spatial differentiation sharpens the boundaries of filaments and clusters. The bottom row presents the tanh-transformed HG (HG-tanh) field $\widetilde{\delta}^\sigma_J$, where the tanh compression reduces extreme values, further enhancing the visibility of intermediate-density structures like connecting filaments. The HG operation thus acts as a gradient-based structural enhancer, while the tanh transformation compresses the dynamic range to reveal a richer, more connected cosmic web.

Crucially, this transformation increases sensitivity to local density gradients while reducing dependence on absolute density values. In the original HG field, cluster and void regions dominate the contrast scale; after tanh transformation, structures across the entire density range contribute more evenly to the visual representation. The resulting enhancement of filamentary patterns and void boundaries is particularly valuable, as these morphological features carry distinct cosmological information relevant to structure-sensitive parameters such as $h$, $M_\nu$, $w$, and $\Omega_b$.

Our results demonstrate that the tanh transformation consistently improves constraints on structure-sensitive parameters ($h$, $M_\nu$, $w$, $\Omega_b$) across all models, as illustrated in Fig.~\ref{fig:constraint_comparison}. In contrast, the effect on amplitude-sensitive parameters ($\Omega_m$, $\sigma_8$) is more model-dependent, with some models showing mild reductions and others achieving modest gains. The combination of the standard power spectrum with derivative-based spectra maintains nearly uniform improvement across all cosmological parameters. Appendix~\ref{app:tanh-notanh-compare} provides a detailed discussion.

\subsection{Power spectrum}
The power spectrum of the HG-tanh magnitude field $\widetilde{\delta}_{J}^\sigma$ is measured using \textsc{Nbodykit} \citep{Hand:2017pqn} to constrain cosmological parameters within the Fisher matrix formalism. For each statistic,  we calculate the monopole up to a maximum wavenumber of $k_{\mathrm{max}} = 0.5\ h\mathrm{Mpc}^{-1}$. We apply a correction for the windowing kernel introduced when interpolating discrete particles to a continuous field \citep{Jing2004CorrectingFT}. Additionally, each power spectrum used in this work is normalized by its mean value. This removes amplitude information and retains only the shape, ensuring consistency with the treatment of real observational data.

The Hermite-Gaussian convolution,  being a differential operator,  introduces a Fourier-space multiplier proportional to $(ik_x)^{n_x}(ik_y)^{n_y}(ik_z)^{n_z} \exp(-k^2)$. This multiplier shapes the resulting power spectrum: the $(i k)^n$ factor suppresses power on very large scales (low $k$),  while the Gaussian term $\exp(-k^2)$ suppresses power on very small scales (high $k$). This combined effect selectively enhances the power spectrum within an intermediate range of scales compared to the standard  power spectrum.

This scale-dependent weighting offers significant advantages. It naturally suppresses noisy small-scale modes and cosmic variance dominated large scale modes,  thereby enhancing the signal within the intermediate,  information-rich range of scales. This effective bandpass filtering can potentially increase the overall Fisher information content for cosmological parameters.

\begin{figure}[t]
    \centering
    \includegraphics[width=0.75\linewidth]{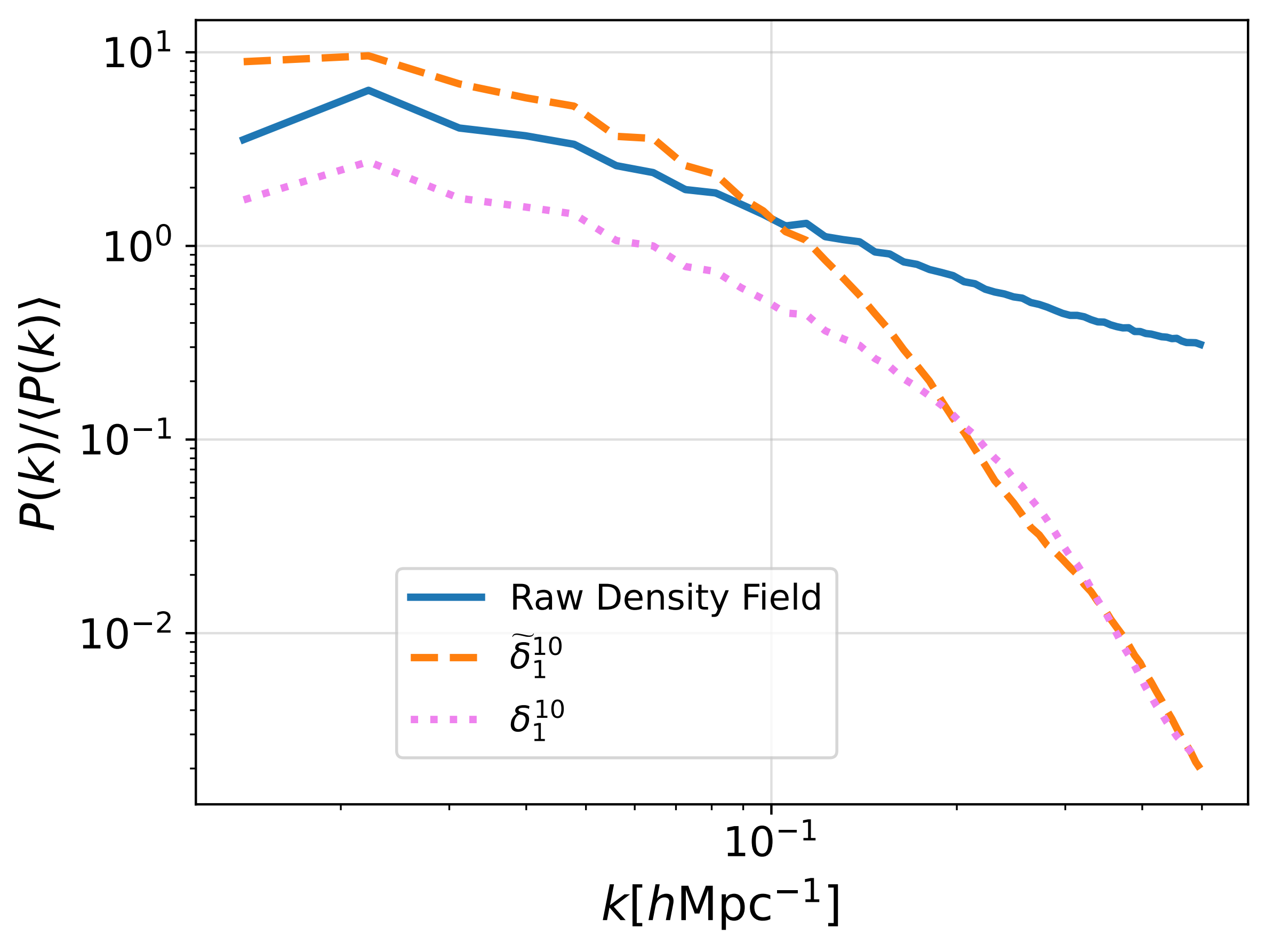}
    \caption{Power spectrum comparison of different field transformations  applied to the Quijote fiducial simulation density field, all using a fixed smoothing scale of $\sigma = 10 h^{-1}\text{Mpc}$. The blue curve shows the raw density field power spectrum $P(k)$.  The purple curve corresponds to the HG convolutional magnitude field $\delta^{10}_{1}$,  while the orange curve shows the HG-tanh magnitude field $\widetilde{\delta}^{10}_{1}(\alpha = 600)$. Each power spectrum is normalized by its mean value $\langle P(k) \rangle$, which is obtained by averaging $P(k)$ over all $k$-bins.}
    \label{fig:ps}
\end{figure}

Fig.~\ref{fig:ps} displays the power spectra of three processed versions of the cosmic density field from the Quijote simulations. Several key features emerge from this comparison:
\begin{itemize}
    \item The raw density field power spectrum (blue curve) exhibits the characteristic monotonic decrease with increasing wavenumber $k$, as expected in $\Lambda$CDM cosmology.
    \item The power spectrum of the HG magnitude field $\delta^{10}_{1}$ (purple) shows a clear, sharp decline near $k\approx0.1\,h\,\mathrm{Mpc}^{-1}$. This feature directly reflects the Gaussian damping factor $\exp(-k^2)$ introduced by the convolution, which effectively suppresses power on scales smaller than the smoothing length, thereby filtering out noise-dominated small-scale modes.
    \item After applying the tanh nonlinearity, the spectrum of $\widetilde{\delta}^{10}_{1}$ (orange) exhibits a notable enhancement on intermediate and large scales (low $k$) compared to the pre-tanh HG field. This enhancement stems from the compressive nature of the $\tanh$ function: by mapping extreme field values toward $\pm 1$, it reduces the dynamic range and suppresses outliers that contribute disproportionately to high-$k$ power. Consequently, the relative weight of coherent, large-scale fluctuations is increased, which amplifies the power on scales where the density field varies smoothly. This processing preserves—and even accentuates—the signal in the scale range most informative for cosmology, while further attenuating noise on very small scales.
\end{itemize}
Together, the HG convolution acts as a scale-dependent bandpass filter, and the subsequent tanh transformation reshapes the filtered spectrum by boosting the large-scale coherent signal relative to the small-scale noise, yielding a final observable that is particularly sensitive to cosmological parameters governing structure morphology.

\subsection{Fisher formalism}

We quantify the information content and cosmological parameter constraints from both the standard and the HG-tanh magnitude field power spectra using the Fisher matrix formalism. For a given set of parameters $\boldsymbol{\theta}$,  the Fisher matrix elements are given by:

\begin{equation}\label{eq:fisher}
F_{ij} = \sum_{\alpha\beta}\frac{\partial S_\alpha}{\partial\theta_i}C^{-1}_{\alpha\beta}\frac{\partial S_\beta}{\partial \theta_j}
\end{equation}
where $S = \{S_0, S_1, ...\}$ is the data vector that can be a single statistic or the concatenation of many of them measured at different wavenumbers $k$,  $\theta$ is the cosmological parameters, and $C$ is the covariance matrix. The inverse of the Fisher matrix,  $F^{-1}$,  provides a lower-bound estimate of the parameter covariance matrix via the Cramér-Rao inequality. This allows us to forecast the constraining power,  $\sigma(\theta_i) \geq \sqrt{(F^{-1})_{ii}}$,  for each cosmological parameter.

\begin{figure}[htbp]
    \centering
    \includegraphics[width=0.7\linewidth]{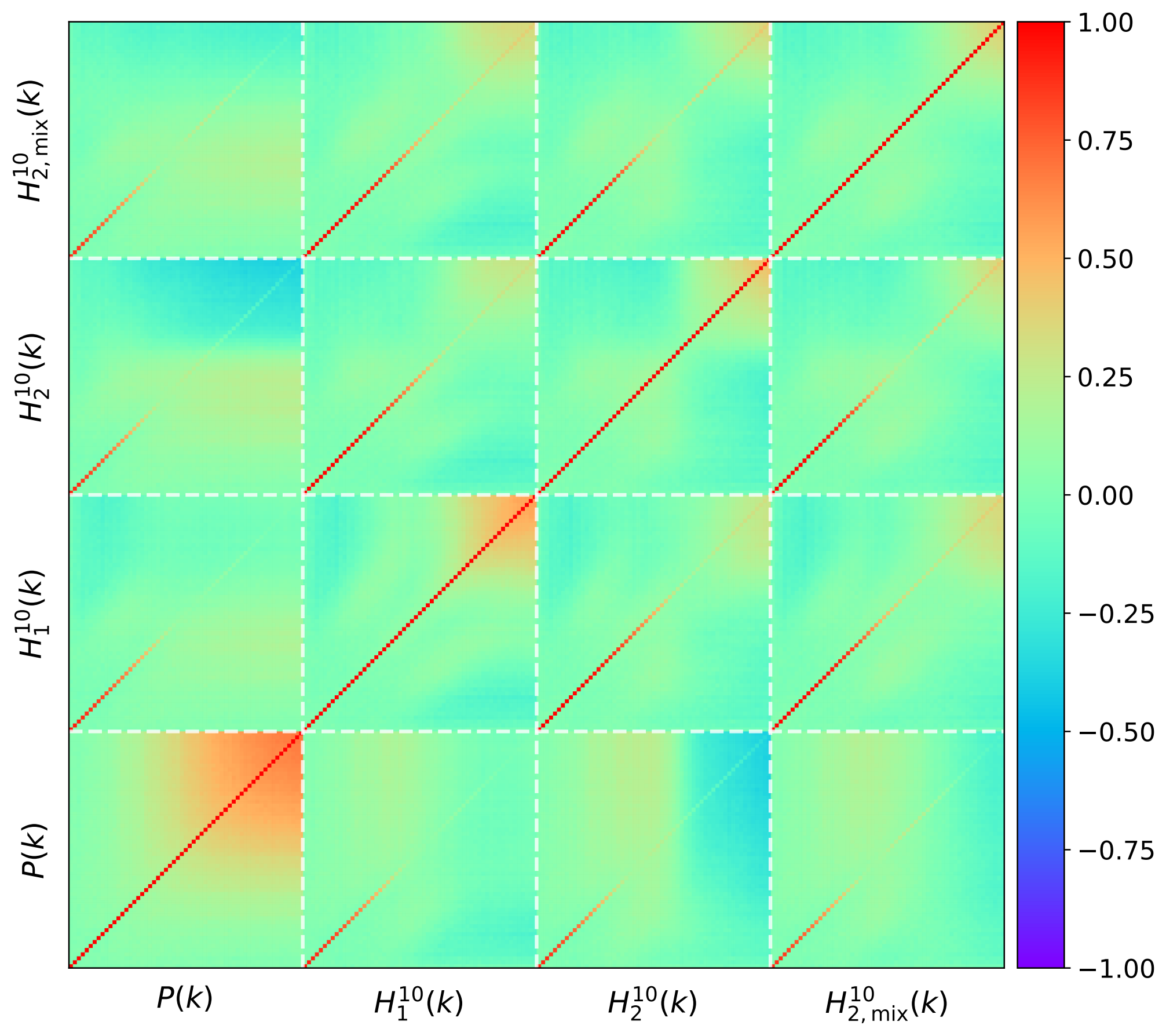}
    \caption{
        Correlation matrices comparing the power spectrum of the raw density field $P(k)$ with those of the HG-tanh magnitude fields. The notation $H^\sigma_{J}(k)$ denotes the power spectrum of the  field $\widetilde{\delta}^\sigma_{J}$. The analysis covers $k < 0.5~h \mathrm{Mpc}^{-1}$ with 59 $k$-bins. The HG-tanh magnitude fields' spectra exhibit enhanced diagonal dominance compared to the standard power spectrum,  particularly on small scales,  indicating reduced mode coupling effects, which allows for more information to be
extracted from small scales.
    }
    \label{fig:corre_matrix}
\end{figure}

The covariance matrix $C$ of different statistics is measured as
\begin{equation}
    \label{eq:cov}
    C_{\alpha\beta} = \langle(S_\alpha- \bar{S_\alpha})(S_\beta- \bar{S_\beta}) \rangle, 
\end{equation}
where $\langle ... \rangle$ indicates the average over different realizations. $\bar{S_i} =\langle S_i\rangle$,  and $S$ is the data vector containing one or
multiple concatenated statistics evaluated at various wavenumbers $k$. We use 10, 000 N-body simulations in fiducial cosmology to compute the covariance.
Fig. \ref{fig:corre_matrix} shows  the correlation matrix ($C_{\alpha\beta}/\sqrt{C_{\alpha\alpha}C_{\beta\beta}}$) for  data vector $\vec{S} = \{P(k), H^{10}_{1}(k), H^{10}_{2}(k), H^{10}_{2,\text{mix}}(k)\}$ (we denote the power sepctrum of the HG-tanh magnitude field  $\widetilde{\delta}^\sigma_{J}$ as $H^\sigma_{J}(k)$ ), All statistics are considered up to a maximum wavenumber $k_\mathrm{max} = 0.5 h \mathrm{Mpc}^{-1}$. We can see that the auto-correlation of $H^{10}_{J}(k)$ display significantly stronger diagonal dominance than the standard power spectrum $P(k)$. This enhanced diagonal structure is most pronounced at small scales ($k > 0.1~h\mathrm{Mpc}^{-1}$),  where nonlinear mode coupling is typically strongest in the standard power spectrum.

We validate the convergence of our covariance matrix estimation in Section~\ref{app:cov_convergence} by systematically testing the stability of parameter uncertainties as the number of mocks increases. The results demonstrate excellent convergence properties,  with all parameter uncertainties stabilizing effectively for $N_{\rm cov}>8000$. This confirms that our covariance matrices are robust and provide a reliable foundation for the cosmological constraints presented in this work.

We need compute the partial derivatives of Eq.\ref{eq:fisher}. For $\Omega_m,  \Omega_b,  h,  n_s,  \sigma_8$ and $w$ we compute
partial derivatives as
\begin{equation}
    \label{eq:deri}
    \frac{\partial \Vec{S}}{\partial \theta} \simeq \frac{\Vec{S}(\theta+d\theta) - \Vec{S}(\theta - d\theta)}{2d\theta}+\mathcal{O}(d\theta^2)
\end{equation}
where $\Vec{S}$ is the considered statistic, and $\theta$ is the cosmological parameter. We thus need to evaluate the statistic
on simulations where only the considered parameter is
varied above and below its fiducial value. The
derivatives w.r.t.\ $M_\nu$ and $w$ have been computed using simulations
with Zel'dovich initial conditions. For $M_\nu$,  instead of Eq.\ \ref{eq:deri},  we use Eq.\ \ref{eq:mnu} since the second
term in the numerator will correspond to a Universe with
negative neutrino masses (see details in the section 2.5 of \citep{Villaescusa-Navarro:2019bje}).
\begin{equation}
    \label{eq:mnu}
    \frac{\partial \vec{S}}{\partial M_\nu} \simeq \frac{-2\vec{S}(2dM_\nu) + 4\vec{S}(dM_\nu) - 3\vec{S}(M_\nu = 0)}{2dM_\nu}+ \mathcal{O}(dM_\nu^2)
\end{equation}
where $dM_\nu$ takes $ 0.1\, \mathrm{eV}$. As mentioned in Section~\ref{sec:simu},  the simulations in massive neutrino cosmologies employ Zel'dovich initial conditions,  while the others use second-order Lagrangian perturbation theory (2LPT) initial conditions. To compute the derivatives in Eq.~\ref{eq:mnu},  the statistics must be measured from both massive and massless neutrino cosmologies. We therefore use simulations with Zel'dovich initial conditions in the fiducial cosmology for these derivative calculations. This approach avoids introducing spurious signals that could arise from differences in the initial condition generation methods \citep{Villaescusa-Navarro:2019bje}. For parameter $w$,  however,  the simulations perturbed above and below its value are based on the same Zel'dovich initial conditions. Consequently,  Eq. \ref{eq:deri} remains valid for it.

\begin{figure}
    \centering
    \includegraphics[width=1\linewidth]{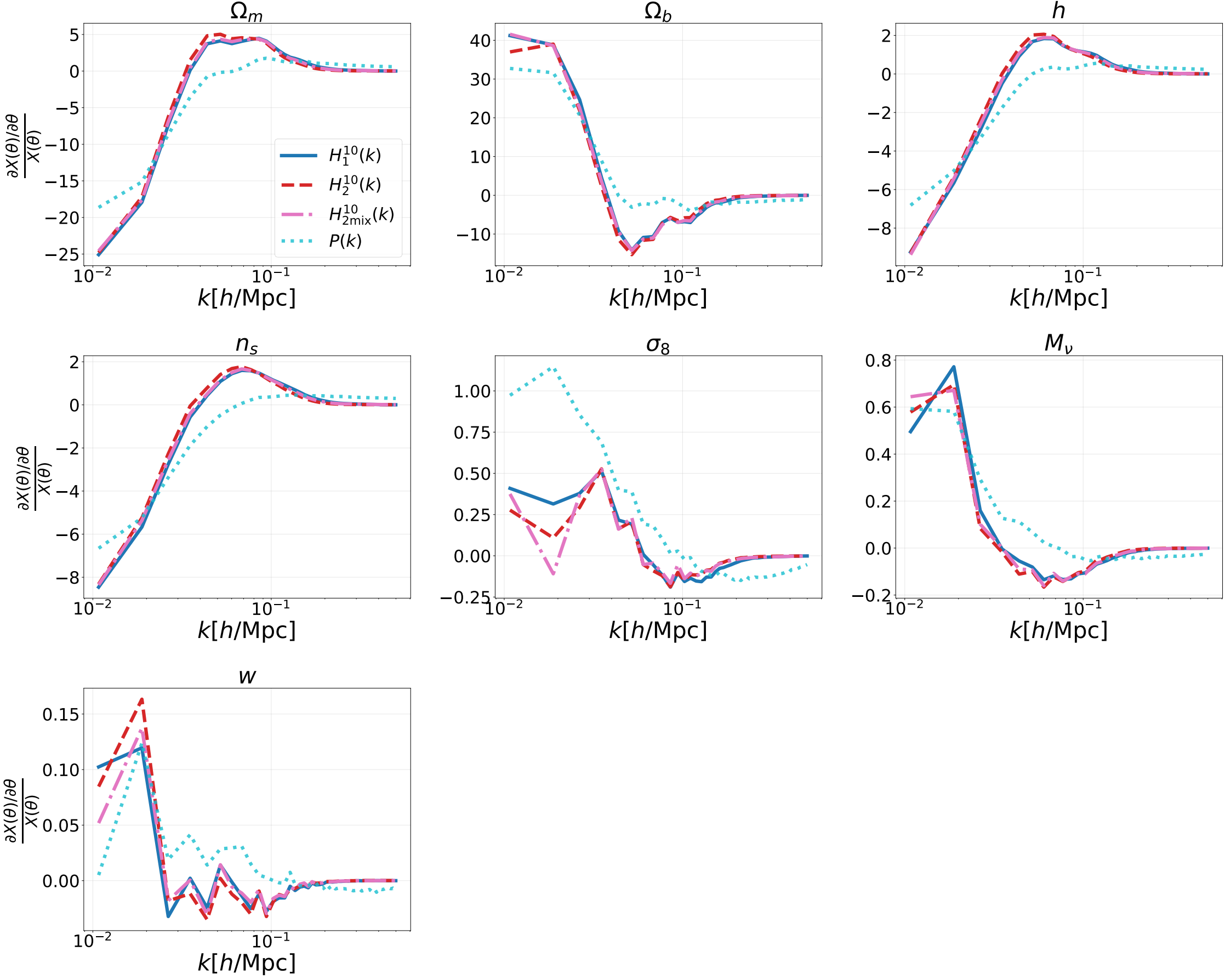}
    \caption{ Numerical logarithmic derivatives with respect to cosmological parameters of the standard power sepctrum $P(k)$ and the transformed field's power spectrum $H^{\sigma}_J(k)$. We show one representative component for each order of the Hermite convolution,  for the smoothing scale $\sigma = 10h^{-1}\mathrm{Mpc}$ case.}
    \label{fig:derivatives}
\end{figure}

Fig. \ref{fig:derivatives} presents a systematic analysis of the sensitivity of both the standard matter power spectrum $P(k)$ and its HG-tanh counterparts to variations in seven key cosmological parameters. The normalized derivatives $\frac{\partial X(\theta)/\partial\theta}{X(\theta)}$ ($X(\theta)$ stands for different power sepctrum) are computed for the standard power spectrum (dotted cyan line) and three representative HG-tanh spectra: the first-order term $H^{10}_{1}(k)$ (solid blue),  second-order term $H^{10}_{2}(k)$ (dashed red),  and the mixed second-order term $H^{10}_{2,\text{mix}}(k)$ (dash-dotted magenta).
The derivatives with respect to each parameter are computed as the mean of 500 simulations. As shown in Figure \ref{fig:derivatives}, the derivatives of the HG-tanh power spectrum at scales $k>0.1\,h\,\mathrm{Mpc}^{-1}$ are similar to those of the standard $P(k)$. In contrast, significant differences appear at $k<0.1\,h\,\mathrm{Mpc}^{-1}$, indicating that the HG-tanh transformation enhances sensitivity to cosmological variations on these larger scales. This provides a complementary approach to standard power spectrum analysis. The power spectra from different orders of HG-tanh fields exhibit similar behavior, although notable differences are observed for parameters $\sigma_8$, $M_\nu$, and $w$ on large scales. We test the convergence of these derivatives in Section~\ref{app:deriv_convergence} by examining the stability of the Fisher analysis results when varying the number of mocks used to compute the derivatives.  Our derivative calculation shows encouraging convergence, with key parameters like $\Omega_m$, $\sigma_8$, and $n_s$ stable at 10\% precision for $N_{\mathrm{der}} > 450$. This supports the robustness of the current Fisher forecasts.

\section{Results and Analysis}
\label{sec:results}
In this section, we present the Fisher forecast $1\sigma$ uncertainties for cosmological parameters using different statistics or their combinations. These include:

\begin{itemize}
    \item The single-order HG-tanh field power spectrum $H_J^{10}(k)$ at a fixed smoothing scale $\sigma = 10\, h^{-1}\mathrm{Mpc}$;
    \item The combined first-order power spectra at different smoothing scales: $H_1^{6}(k) + H_1^{10}(k) + H_1^{15}(k)$;
    \item The combined spectra of different orders at a fixed smoothing scale $\sigma = 10\, h^{-1}\mathrm{Mpc}$, namely $H_1^{10}(k) + H_2^{10}(k) + H_{2,\text{mix}}^{10}(k)$;
    \item The combination of the standard power spectrum and one HG-tanh spectrum: $P(k) + H^{10}_1(k)$;
    \item The combination of all nine HG-tanh field power spectra considered in this work (three orders and three smoothing scales), labeled as ``All $H(k)$'';
    \item The combination of the standard power spectrum and all HG-tanh power spectra, labeled as``$P(k)$ + All $H(k)$''.
\end{itemize}
\begin{table}
\centering
\footnotesize
\setlength{\tabcolsep}{2pt}  
\caption{Fisher forecast $1\sigma$ error of cosmological parameters from different observables,  when including the wavenumber with $k<k_{\mathrm{max}}=0.5\,h\mathrm{Mpc}^{-1}$.The observables $H^\sigma_1(k)$,  $H^\sigma_2(k)$ and $H^\sigma_{2,\mathrm{mix}}(k)$ represent the spectrum of HG-tanh fields $\widetilde{\delta}^{\sigma}_{1}$, $\widetilde{\delta}^{\sigma}_{2}$ and  $\widetilde{\delta}^{\sigma}_{2,\text{mix}}$  respectively. "All $H(k)$" refers to the power spectrum obtained by concatenating the spectra of all HG-tanh fields: the first- $H^\sigma_1(k)$, second- $H^\sigma_2(k)$, and mixed second-order $H^\sigma_{2,\text{mix}}(k)$, each evaluated at the three smoothing scales $\sigma = 6, 10,$ and $15$ $h^{-1}$Mpc. We employ leave-one-out cross-validation to compute the reduced $\chi^2_\nu$: for each fiducial cosmology realization, the covariance matrix is constructed from the remaining $9999$ realizations, and $\chi^2_\nu$ is calculated for the excluded realization. This procedure is repeated 1500 times. In the bottom two rows, we present the mean and standard deviation of the observed ratio ${\chi^2_{\nu,\mathrm{Obs}}}$ and the theoretical  ratio ${\chi^2_{\nu,\mathrm{Theory}}}$. The histogram distribution is compared with the theoretical expectation $\mathcal{N}\bigl(1,\sqrt{2/\nu}\bigr)$ in Appendix~\ref{app:chi2dof_hist}.}
\begin{tabular}{|c|c|c|c|c|c|c|c|c|c|c|}
\hline
 & $P(k)$ & $H_{1}^{10}(k)$ & $H_{2}^{10}(k)$ & $H_{2,\mathrm{mix}}^{10}(k)$ & \begin{tabular}{@{}c@{}}$H_{1}^{6}(k)$\\$+H_{1}^{10}(k)$\\$+H_{1}^{15}(k)$\end{tabular} & \begin{tabular}{@{}c@{}}$P(k)$\\$+H_{1}^{10}(k)$\end{tabular} & \begin{tabular}{@{}c@{}}$H_{1}^{10}(k)$\\$+H_{2}^{10}(k)$\\$+H_{2,\mathrm{mix}}^{10}(k)$\end{tabular} & \begin{tabular}{@{}c@{}}All\\$H(k)$\end{tabular} & \begin{tabular}{@{}c@{}}$P(k)+$\\All $H(k)$\end{tabular} & \begin{tabular}{@{}c@{}}$P(k)$/\\$[P(k)+$\\All $H(k)]$\end{tabular} \\
\hline
$\Omega_{m}$ & 0.0288 & 0.0387 & 0.0308 & 0.0347 & 0.0228 & 0.0213 & 0.0227 & 0.0163 & 0.0143 & 2.0 \\
$\Omega_{b}$ & 0.0129 & 0.0087 & 0.0091 & 0.0100 & 0.0057 & 0.0075 & 0.0055 & 0.0031 & 0.0031 & 4.2 \\
$h$ & 0.1546 & 0.1048 & 0.1122 & 0.1208 & 0.0682 & 0.0857 & 0.0691 & 0.0444 & 0.0431 & 3.6 \\
$n_{s}$ & 0.0918 & 0.1304 & 0.0899 & 0.1046 & 0.0584 & 0.0607 & 0.0594 & 0.0348 & 0.0321 & 2.9 \\
$\sigma_{8}$ & 0.0452 & 0.0725 & 0.0495 & 0.0916 & 0.0381 & 0.0342 & 0.0301 & 0.0153 & 0.0138 & 3.3 \\
$M_{\nu}(\mathrm{eV})$ & 0.3860 & 0.2748 & 0.2723 & 0.2673 & 0.1656 & 0.2129 & 0.1477 & 0.0842 & 0.0809 & 4.8 \\
$w$ & 0.7568 & 0.4488 & 0.5013 & 0.4643 & 0.2547 & 0.3535 & 0.2609 & 0.1481 & 0.1431 & 5.3 \\
\hline
\multirow{2}{*}{${\chi^2_{\nu,\mathrm{Obs}}}$} & 1.004 & 1.000 & 1.002 & 0.991 & 1.001 & 0.999 & 1.004 & 1.000 & 1.000 & - \\
 & $\pm$0.187 & $\pm$0.187 & $\pm$0.183 & $\pm$0.179 & $\pm$0.106 & $\pm$0.133 & $\pm$0.106 & $\pm$0.06 & $\pm$0.058 & - \\
\hline
\multirow{2}{*}{${\chi^2_{\nu,\mathrm{Theory}}}$} & 1 & 1 & 1 & 1 & 1 & 1 & 1 & 1 & 1 & - \\
 & $\pm$0.184 & $\pm$0.184 & $\pm$0.184 & $\pm$0.184 & $\pm$0.106 & $\pm$0.130 & $\pm$0.106 & $\pm$0.061 & $\pm$0.058 & - \\
\hline
\end{tabular}
\label{tab:fisher_results}
\end{table}

The Fisher forecast $1\sigma$  constraints in Table~\ref{tab:fisher_results} demonstrate that the power spectrum of HG-tanh magnitude fields extracts complementary information compared to the standard $P(k)$. The analysis reveals distinct sensitivity patterns for structure-sensitive parameters ($h$, $M_{\nu}$, $w$, $\Omega_{b}$) versus amplitude-sensitive parameters ($\Omega_{m}$, $\sigma_{8}$).

For a fixed smoothing scale $\sigma=10$, individual HG-tanh statistics $H^{10}_{1}(k)$, $H^{10}_{2}(k)$ and $H^{10}_{2,\mathrm{mix}}(k)$, consistently improve constraints on structure-sensitive parameters relative to $P(k)$, but yield mixed results for amplitude-sensitive ones. Specifically, $H^{10}_{1}(k)$ improves constraints on $\Omega_{b}$, $h$, $M_{\nu}$, and $w$ by 33\%, 32\%, 29\%, and 41\%, respectively, while slightly degrading $\Omega_{m}$ and $\sigma_{8}$ by 34\% and 60\%. The spectral index $n_{s}$ shows minimal change. Similar patterns hold for $H^{10}_{2}(k)$ and $H^{10}_{2,\mathrm{mix}}(k)$, confirming that individual HG-tanh statistics trade amplitude sensitivity for enhanced morphological information.

Combining the standard $P(k)$ with $H^{10}_{1}(k)$ yields improvement for every parameter, with gains ranging from 26\% ($\Omega_{m}$) to 53\% ($w$). This combined improvement indicates that the HG-tanh field and $P(k)$ capture complementary aspects of the cosmological information, with the HG-tanh statistics adding new constraints beyond what the standard power spectrum can provide.

Systematic combination of spectra across orders or scales yields progressively stronger constraints. Combining the three orders  $H^{10}_{1}(k)+H^{10}_{2}(k)+H^{10}_{2,\mathrm{mix}}(k)$ at $\sigma=10h^{-1}\mathrm{Mpc}$ improves all parameters by 41--81\% relative to $P(k)$. Combining three smoothing scales for the first-order field $H^{6}(k)_{1}+H^{10}_{1}(k)+H^{15}_{1}(k)$ achieves even larger gains of 43--81\%. The most comprehensive combination---all nine HG-tanh fields (three orders $\times$ three scales)---produces stronger overall improvement. Adding $P(k)$ to this full set yields a further modest tightening, with factors of 2.0--5.3$\times$ relative to $P(k)$ alone. This result suggests that, in terms of the Fisher information captured, the HG-tanh spectra largely subsume the information content of the standard power spectrum.

The $\chi^{2}_\nu$ diagnostics reported in Table~\ref{tab:fisher_results} validate the accuracy of our covariance estimation. For all models, the observed means (ranging from 0.991 to 1.004) align closely with the theoretical expectation of unity, and the standard deviations are consistent with the Gaussian prediction $\mathcal{N}(1,\sqrt{2/\nu})$, where $\nu$ represents the d.o.f. of the ccorresponding statistic. The full distribution of $\chi^{2}_\nu$, shown in Appendix~\ref{app:chi2dof_hist}, further confirms the statistical reliability of the Fisher matrix results presented in Table~\ref{tab:fisher_results}.

\begin{figure}
\centering
\includegraphics[width=\textwidth]{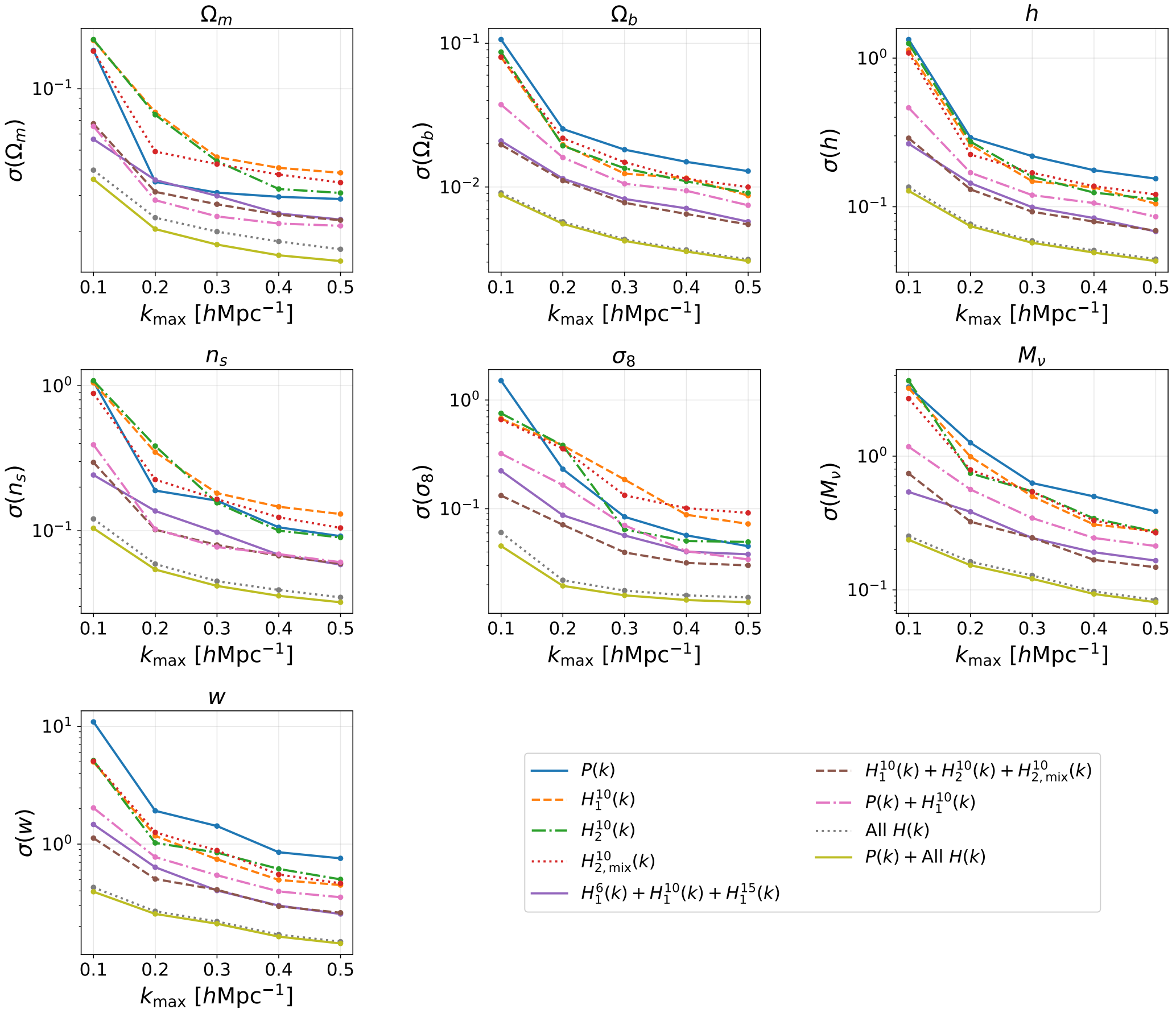}
\caption{
Parameter constraints as a function of maximum wavenumber $k_{\mathrm{max}}$ for different statistical observables. 
The Fisher forecast $1\sigma$ errors are shown for seven cosmological parameters: 
$\Omega_m$,  $\Omega_b$,  $h$,  $n_s$,  $\sigma_8$,  $M_\nu$,  and $w$. 
9 different statistical approaches are compared: the standard power spectrum $P(k)$; 
first-order $H_1^{10}(k)$,  second-order $H_2^{10}(k)$,  and mixed second-order $H_{2,\mathrm{mix}}^{10}(k)$ Hermite convolutional power spectra; 
and their various combinations. 
The constraints generally improve with increasing $k_{\mathrm{max}}$ as more Fourier modes are included,  
with the combined approaches showing significant enhancement over individual statistics.}
\label{fig:parameter_errors_kmax}
\end{figure}

Figure~\ref{fig:parameter_errors_kmax} presents the evolution of cosmological parameter constraints as a function of the maximum wavenumber $k_{\mathrm{max}}$ for different statistical approaches. The key observations are as follows:
First,  for the $\Omega_m$ parameter,  the standard power spectrum $P(k)$ shows saturation in constraining power when $k_{\mathrm{max}} > 0.2$,  while the HG-tanh field power spectra continue to improve with increasing wavenumber. This clearly demonstrates that the HG-tanh field approach extracts additional cosmological information beyond what is accessible to the standard power spectrum.
As expected,  all methods show improved constraints with increasing $k_{\mathrm{max}}$,  since more Fourier modes are included,  providing more information about the density field. This trend is consistent with theoretical predictions.
Overall,  the HG-tanh field power spectra outperform the standard $P(k)$ for most cosmological parameters,  which is consistent with the Fisher matrix analysis results presented in Table~\ref{tab:fisher_results}. This consistency validates the effectiveness of HG-tanh field statistics in extracting cosmological information from large-scale structure data.

\begin{figure}
\centering
\includegraphics[width=\textwidth]{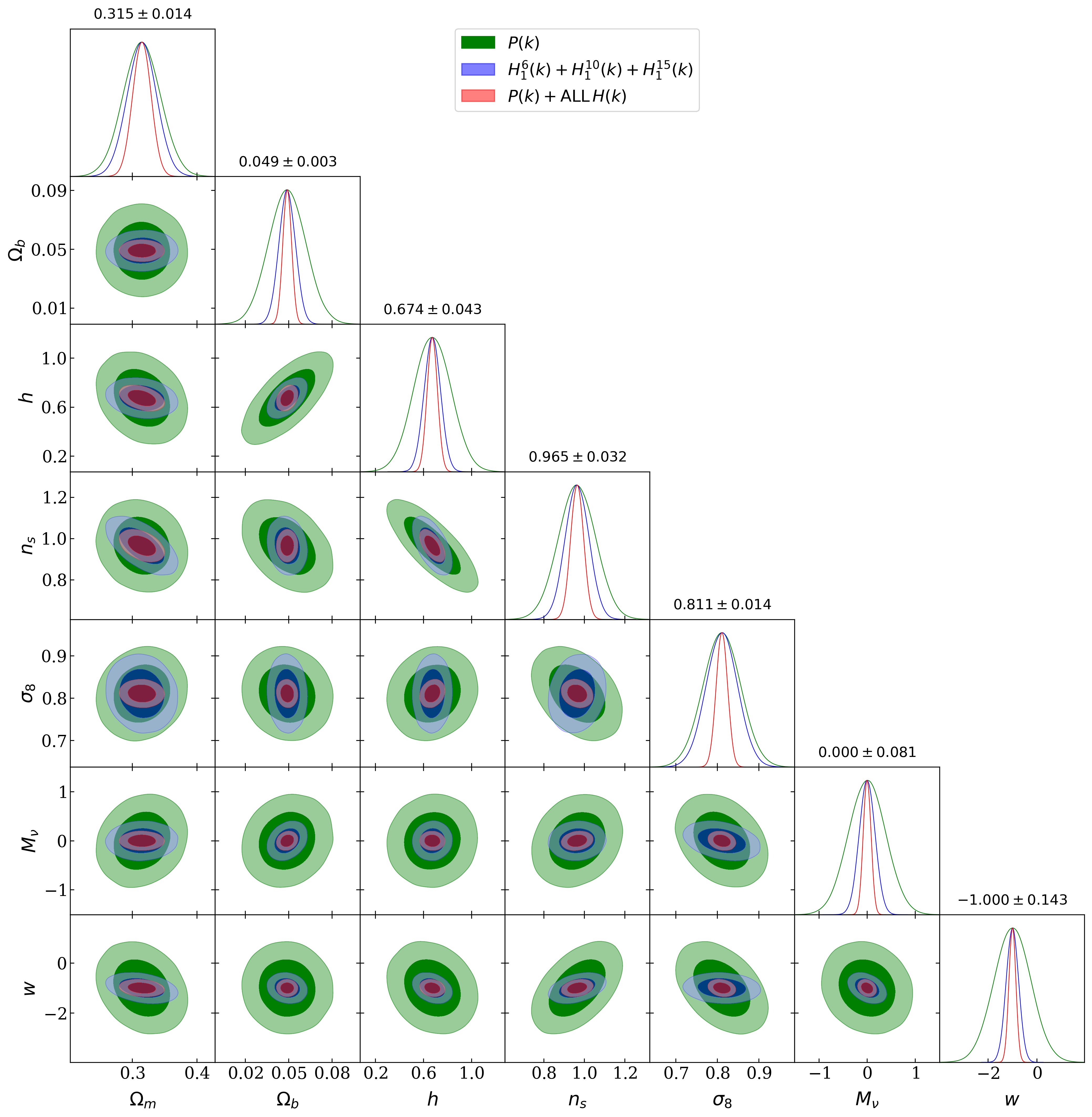}
\caption{
Fisher matrix constraints (darker and lighter shades being the 68\% and 95\% confidence contours) on cosmological
parameters obtained  using the model $P(k)$ (green), the combination $H_{1}^{6}(k) + H_{1}^{10}(k) + H_{1}^{15}(k)$ (blue),and $P(k) + \mathrm{All}\, H(k)$ (red). The constraint shown for each parameter results from the model $P(k) + \mathrm{All}\, H(k)$. The maximum wavenumber for each observable is set to $k_{\mathrm{max}} = 0.5 h\mathrm{Mpc}^{-1}$}.
\label{fig:2dpost}
\end{figure}

Figure~\ref{fig:2dpost} shows the 1$\sigma$ and 2$\sigma$ constraints on seven cosmological parameters obtained from Fisher analysis of the standard power spectrum $P(k)$ (green contours), the first order HG-tanh field combination $H_1^{6}(k) + H_1^{10}(k) + H_1^{15}(k)$ (blue), and the full combination $P(k) + \mathrm{All}\, H(k)$ (red). The combination of first-order fields at different scales significantly tightens the constraints on all parameters except  $\Omega_m$. The joint analysis breaks several key degeneracies, notably between $\sigma_8$ and $M_\nu$ , between $n_s$ and $\sigma_8$, and between $w$ and $\sigma_8$, resulting in the tightest constraints across all parameters (red). The derivative nature of the HG-tanh transform enhances its sensitivity to the sharp features imprinted by massive neutrinos and dark energy, complementing the broadband shape information captured by $P(k)$. This multi-scale, gradient-based approach extracts complementary cosmological information from the large-scale structure, demonstrating the value of combining traditional and Hermite-Gaussian statistics for precision cosmology.

\section{Conclusion}
\label{sec:conclusion}

In this work,  we have developed and validated a comprehensive framework for enhancing cosmological parameter constraints through Hermite-Gaussian convolution and tanh transformations of the cosmic density field. Our approach addresses three fundamental aspects of large-scale structure analysis: derivative field computation,  cosmic web enhancement,  and cosmological information extraction.

The Hermite-Gaussian convolutional method provides a robust mathematical foundation for computing arbitrary-order derivatives of the density field while simultaneously suppressing small-scale noise through the inherent Gaussian cutoff. This dual functionality ensures that the derived statistics maintain both mathematical rigor and physical relevance. Furthermore, multi-scale information can be captured by varying the size of the convolution kernel. The application of the hyperbolic tangent transformation to these convolved fields has proven particularly effective in enhancing the visibility of cosmic web structures,  making voids,  clusters,  and filaments more distinguishable in the transformed fields.

Our comprehensive Fisher analysis,  based on the Quijote simulations,  demonstrates the remarkable effectiveness of this approach. The key findings can be summarized as follows:

First,  the power spectra of Hermite-convolved fields contain complementary cosmological information to the standard power spectrum. This is evidenced by the significant improvement in parameter constraints when combining both statistics,  with the combination $P(k) + H_1^{10}(k) $ achieving significant improvements compared with the standard power spectrum $P(k)$.

Second,  the multi-scale approach proves particularly powerful,  with different smoothing scales ($\sigma = 6,  10,  15$ $h^{-1}$Mpc) capturing distinct aspects of cosmological information. Smaller scales probe non-linear regime effects,  while larger scales sample linear regime information,  together providing a more complete picture of structure formation.

Third,  the method shows exceptional sensitivity to parameters that are traditionally challenging to constrain,  such as neutrino mass and dark energy equation of state. The 3.9$\times$ and 4.3$\times$ improvements for $M_\nu$ and $w$,  respectively,  highlight the method's potential for addressing fundamental questions in cosmology.

The convergence tests presented in the appendices confirm the robustness of our covariance matrix estimation,  while also identifying areas for future refinement in numerical derivative computation. Although the current implementation already provides substantial improvements,  there remains potential for further enhancement through optimized derivative calculation techniques.

In conclusion,  the Hermite-Gaussian convolutional  represents a novel approach in LSS analysis,  providing a mathematically effective and observationally powerful tool for extracting multi-scale cosmological information from the complex patterns of the cosmic web. The  improvements in parameter constraints demonstrated in this work suggest that this method will play an important role in maximizing the scientific return from next-generation cosmological surveys.

\appendix
\section{Parameter Constraints Using the Power Spectra of Convolutional Fields and Their Tanh-Transformed Counterparts}
\label{app:tanh-notanh-compare}
\begin{figure}[htbp]
\centering
\includegraphics[width=0.8\textwidth]{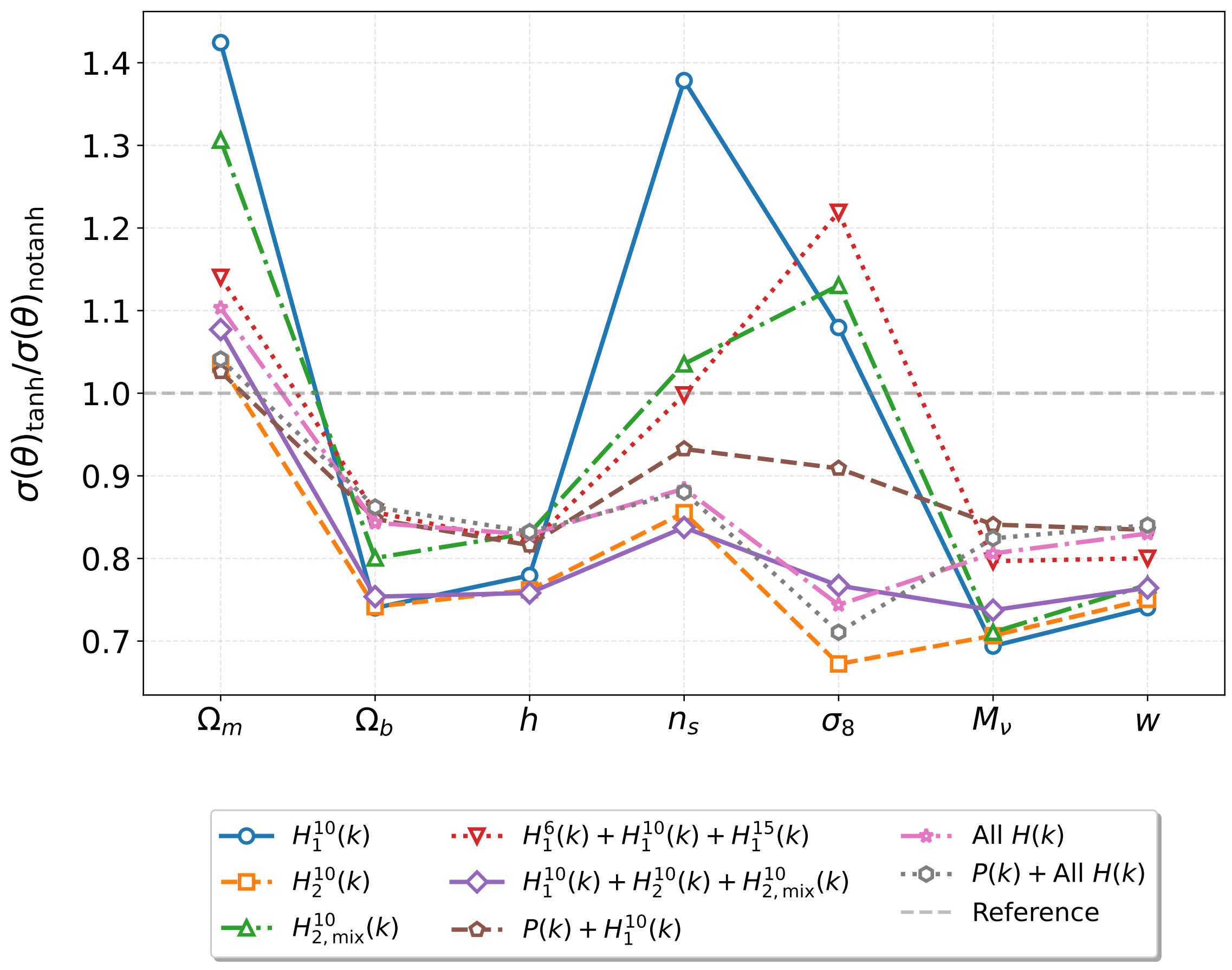}
\caption{
Comparison of parameter constraints between the HG field power spectrum and the HG-tanh field power spectrum across all 9 models. The $y$-axis shows the ratio of the $1\sigma$ errors (HG-tanh to HG), with values below unity indicating improved constraints from the tanh transformation.
}
\label{fig:constraint_comparison}
\end{figure}

The comparative analysis reveals distinct patterns in parameter constraint improvements when applying the tanh nonlinear transformation to convolutional fields.

Figure~\ref{fig:constraint_comparison} demonstrates that the tanh transformation provides selective enhancement of cosmological parameter constraints:

\begin{itemize}
    \item \textbf{Structure-sensitive parameters} ($h$, $M_\nu$, $w$, $\Omega_b$) show consistent improvement across all constraint scenarios. This enhancement stems from the tanh function's ability to compresses extreme density contrasts and enhances the visibility of cosmic web structures, making subtle topological features more pronounced in the density field.

    \item \textbf{Amplitude-sensitive parameters} ($\Omega_m$, $\sigma_8$) exhibit a different response to the tanh transformation. For $\Omega_m$, the error ratio $\sigma(\theta)_{\text{tanh}} / \sigma(\theta)_{\text{no-tanh}}$ is consistently greater than unity across all nine models, indicating that the tanh operation systematically degrades the constraint on the total matter density. This can be understood as a consequence of the compressive nonlinearity: by mapping extreme density contrasts toward $\pm 1$, the $\tanh$ function suppresses the dynamic range of the field. This suppression disproportionately affects the higher-order statistical moments that are crucial for precisely measuring the overall matter amplitude, to which $\Omega_m$ is particularly sensitive. In essence, while tanh enhances the visibility of cosmic web structures, it “flattens” the full amplitude distribution, thereby losing some of the fine-grained information about the absolute mass density. The effect on $\sigma_8$ is more model-dependent. In some combinations, a mild improvement is visible, whereas in others the constraint is slightly worse. This variability suggests that the impact of tanh on the clustering amplitude is contingent on which other parameters are simultaneously constrained and how the corresponding degeneracy directions are oriented in the full parameter space. The fact that $\sigma_8$ does not show a uniform degradation like $\Omega_m$ may indicate that the amplitude of \textit{rms} fluctuations is partially encoded in the spatial arrangement of structures—information that the tanh-enhanced morphology can still capture to some degree.
\end{itemize}
Overall, the tanh transformation trades amplitude sensitivity for enhanced morphological contrast. This trade-off is beneficial for structure-sensitive parameters ($h$, $M_\nu$, $w$, $\Omega_b$) but comes at a cost for parameters that rely heavily on the absolute amplitude of the density field, most clearly seen in the case of $\Omega_m$. The spectral index $n_s$, which governs the tilt of the primordial power spectrum, benefits moderately from the tanh transformation. The spectral index $n_s$ occupies an intermediate position: its constraints are consistently and moderately improved by the tanh operation, with error ratios clustering just below unity. This pattern aligns with $n_s$'s role as a shape parameter—it is sensitive to the relative amplitude of fluctuations across scales rather than their absolute magnitude. The tanh transformation, by enhancing multi-scale contrast, appears to better preserve this scale-relative information while suppressing the absolute amplitude information that is crucial for $\Omega_m$.

\section{Convergence Test of Covariance Matrix Estimation}
\label{app:cov_convergence}

\begin{figure}[htbp]
\centering
\includegraphics[width=0.95\textwidth]{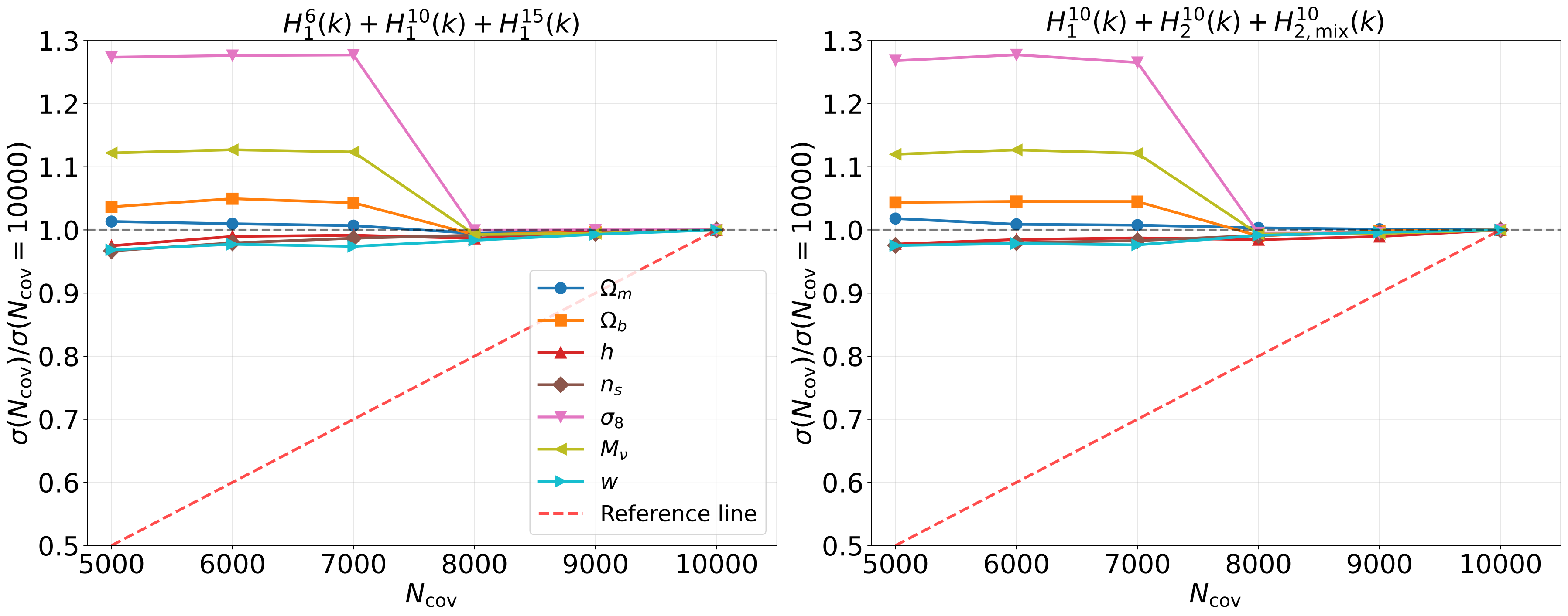}
\caption{Convergence test of covariance matrix estimation. Both panels show the ratio of parameter uncertainties $\sigma(N_{\text{cov}})$ to the reference value $\sigma(N_{\text{cov}}=10000)$ as a function of the number of mock realizations. The left panel corresponds to the data vector $H_{1}^{6}(k)+H_{1}^{10}(k)+H_{1}^{15}(k)$,  while the right panel shows results for $H_{1}^{10}(k)+H_{2}^{10}(k)+H_{2,\text{mix}}^{10}(k)$. The red dashed reference line indicates the variation in number of mocks compare to the full sample. All seven cosmological parameters ($\Omega_m$,  $\Omega_b$,  $h$,  $n_s$,  $\sigma_8$,  $M_\nu$,  $w$) show good convergence behavior as $N_{\text{cov}}$ increases,  with uncertainties stabilizing near the reference value for $N_{\text{cov}} > 8000$.}
\label{fig:app_cov_convergence}
\end{figure}

Figure~\ref{fig:app_cov_convergence} presents a comprehensive convergence analysis of the covariance matrix estimation procedure. We systematically examine how the parameter uncertainties depend on the number of mock realizations used for covariance estimation,  with $N_{\text{cov}}$ ranging from 5, 000 to 10, 000.The uncertainties for all parameters converge satisfactorily for $N_{\text{cov}} > 8000$,  with the ratios stabilizing near unity. This demonstrates that using 10, 000 mock realizations provides well-converged covariance matrices for reliable cosmological parameter constraints.

\section{Convergence Test of Numerical Derivatives}
\label{app:deriv_convergence}

\begin{figure}[htbp]
\centering
\includegraphics[width=0.95\textwidth]{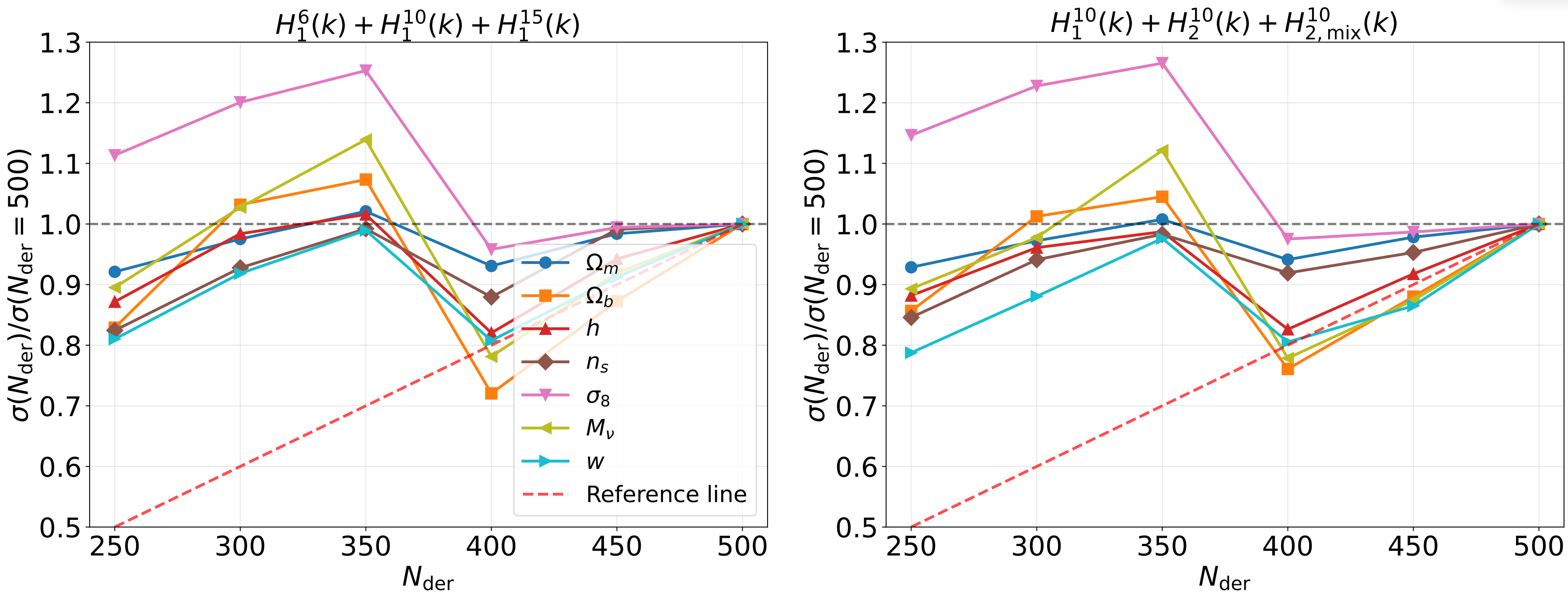}
\caption{Convergence test of numerical derivative estimation. Both panels show the ratio of parameter uncertainties $\sigma(N_{\text{der}})$ to the reference value $\sigma(N_{\text{der}}=500)$ as a function of the number of mock realizations used for derivative computation. The left panel corresponds to the data vector $H_{1}^{6}(k)+H_{1}^{10}(k)+H_{1}^{15}(k)$,  while the right panel shows results for $H_{1}^{10}(k)+H_{2}^{10}(k)+H_{2,\text{mix}}^{10}(k)$.  The overall convergence trend becomes clearly established for $N_{\mathrm{der}} > 450$.}
\label{fig:app_deriv_convergence}
\end{figure}

Figure~\ref{fig:app_deriv_convergence} presents the convergence analysis of numerical derivatives used in the Fisher matrix calculation. Unlike the covariance matrix convergence test,  the derivative convergence shows a more gradual and incomplete convergence pattern. While the parameter uncertainties begin to stabilize for $N_{\text{der}} > 450$,  the curves do not fully flatten,  indicating that the derivative estimation has not reached complete convergence within the tested range of 250-500 mock realizations.

The derivatives demonstrate acceptable convergence for cosmological constraints, with parameters $\Omega_m$, $\sigma_8$, and $n_s$ reaching 10\% precision. The overall convergence trend becomes clearly established for $N_{\mathrm{der}} > 450$, providing confidence in the current Fisher forecasts. Future improvements through increased mock realizations or refined differentiation schemes can further strengthen the robustness of the analysis.

\section[Distribution of chi2/d.o.f. from cross-validation]
        {Distribution of $\chi^2_\nu$ from cross-validation}
\label{app:chi2dof_hist}
\begin{figure}[htbp]
    \centering
    \includegraphics[width=0.95\linewidth]{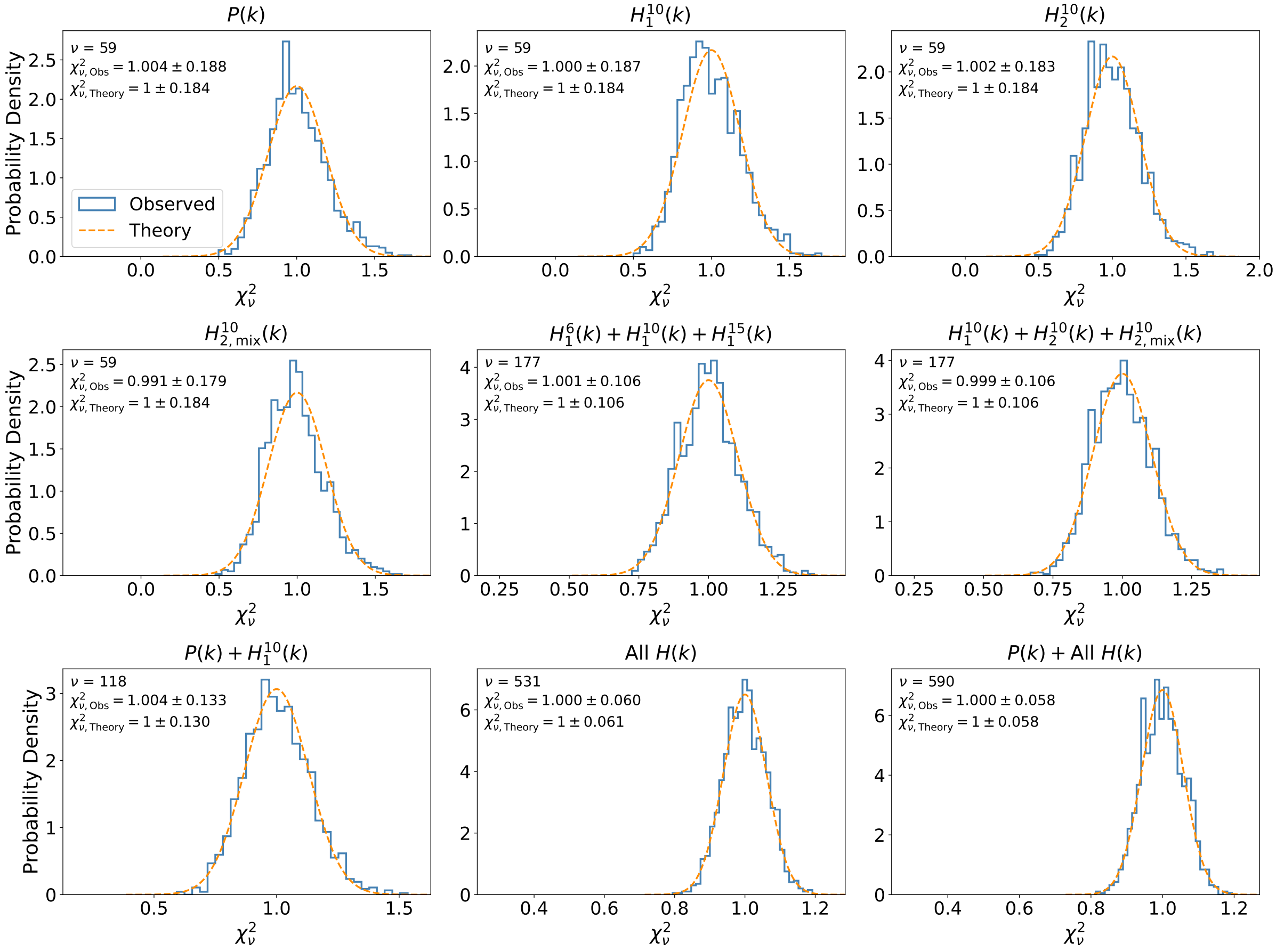}
    \caption{
        Distributions of  $\chi^2_\nu$ for different statistical models, computed via leave-one-out cross-validation. Blue histograms show the observed distributions from 1500 realizations, while orange dashed curves represent the theoretical expectation $\mathcal{N}(1,\sqrt{2/\nu})$. Excellent agreement is found across all model configurations, validating the reliability of the estimated covariance matrices. Observed means (annotated in each panel) consistently lie within one standard deviation of the theoretical value of unity, confirming the statistical accuracy of our covariance estimation procedure.
    }
    \label{fig:chi2dof}
\end{figure}

The leave-one-out cross-validation demonstrates that our covariance matrices are accurately estimated. For each of the 9 model configurations, the distribution of reduced $\chi^2_\nu$ values (Fig.~\ref{fig:chi2dof}) closely matches the theoretical Gaussian distribution $\mathcal{N}(1,\sqrt{2/\nu})$. The observed mean values, ranging from 0.991 to 1.004, consistently agree with the theoretical expectation of unity to within one standard deviation. This agreement across varying degrees of freedom (59 to 590) confirms the statistical reliability of our covariance matrices, ensuring robust parameter constraints in the Fisher matrix analysis.

\section*{ACKNOWLEDGMENTS}
\label{sec:acknowledgments}
This work was supported by the National SKA Program of China (2020SKA0110401,  2020SKA0110402, 
2020SKA0110100),  the National Key R$\&$D Program of
China (2020YFC2201600),  the National Natural Science
Foundation of China (12373005,  12473097,  12073088), 
the China Manned Space Project (CMS-CSST-2021:
A02,  A03,  B01),  and the Guangdong Basic and Applied
Basic Research Foundation (2024A1515012309). We acknowledge the Beijing Super Cloud Center (BSCC) and
Beijing Beilong Super Cloud Computing Co.,  Ltd.
(http://www.blsc.cn/) for providing HPC resources that
substantially supported this study.

\bibliographystyle{raa}
\bibliography{bibtex}

\label{lastpage}

\end{document}